\documentclass[twocolumn,         % Format : preprint, twocolumn
               showpacs,longbibliography,            % Pacs : showpacs, noshowpacs
               showkeys,preprintnumbers,     % Preprint: preprintnumbers,
               aps,                 % Society: ...
               prd,          	    % Journal Style : pra, prb, prc, prd, pre,
               letterpaper,             % Size : a4paper, ...
               superscriptaddress,      % Affiliation (Title) : groupedaddress,
               nofootinbib,         % Footnote: footinbib, nofootinbib
               tightenlines,        % Remove additional spaces in a line
               floats,floatfix      % Floating pictures and tables
               ]{revtex4-1}
\usepackage{float}
\usepackage{dblfloatfix}
\usepackage{physics}
\usepackage{epsf}
\usepackage{graphicx,color}
\usepackage{subfigure}
\usepackage{latexsym}
\usepackage{placeins}
\usepackage{amsmath,amssymb}        
\usepackage[colorlinks=true,linkcolor=blue,citecolor=blue]{hyperref}
\usepackage{mathrsfs}
\usepackage{comment}
\usepackage{soul}
\usepackage{cancel}
\definecolor{purple}{rgb}{0.58,0.0,0.83}
\definecolor{orange}{rgb}{1,0.5,0}
\DeclareSymbolFontAlphabet{\mathrsfs}{rsfs}
\DeclareMathAlphabet{\mathcal}{OMS}{cmsy}{m}{n}

\begin{document}

% -----> TITLE 

\title{Equilibrium Core and Vortex Solutions of Bose Einstein Condensate Dark Matter around a Black Hole}
% ----->   AUTHORS   <-----

\author{Iv\'an  \'Alvarez-Rios}
\email{ivan.alvarez@umich.mx}
\affiliation{Instituto de F\'{\i}sica y Matem\'{a}ticas, Universidad
              Michoacana de San Nicol\'as de Hidalgo. Edificio C-3, Cd.
              Universitaria, 58040 Morelia, Michoac\'{a}n,
              M\'{e}xico.}               

\author{Francisco S. Guzm\'an}
\email{francisco.s.guzman@umich.mx}
\affiliation{Instituto de F\'{\i}sica y Matem\'{a}ticas, Universidad
              Michoacana de San Nicol\'as de Hidalgo. Edificio C-3, Cd.
              Universitaria, 58040 Morelia, Michoac\'{a}n,
              M\'{e}xico.}       
% --->   DATE

\date{\today}

% -----> ABSTRACT

\begin{abstract}
We present the construction of stationary solutions of Bose-Einstein condensate dark matter (BECDM) around a point-like gravitational source representing a black hole. The problem is formulated for general axisymmetric configurations, and we focus on two cases: the ground-state core solution and the first nonzero winding number configuration corresponding to a line vortex solution. The stationary equations are solved using an imaginary-time approach, which enables the construction of families of solutions across a wide range of self-interaction and black hole masses. We analyze the impact of these parameters on the density distribution and on the stability properties of the solutions, assessing stability through the turning point criterion based on the enthalpy functional, which allows us to identify stable and unstable branches along each family of solutions. It has been shown in the past that spherical core solutions act as attractors in the collapse of BECDM around black holes in the non-interacting case ($g=0$), supporting their astrophysical relevance. In the present work, the existence of a maximum mass for configurations with attractive self-interaction ($g<0$) allows us to infer the parameter range in which such solutions may also arise in this regime. Building on this picture, we show that stable vortex solutions of BECDM can also exist in the presence of a black hole, whose stability properties suggest that these configurations may likewise be compatible with physically relevant formation scenarios.
\end{abstract}

% ----->   PACS

% ----->   MAKETITLE   <-----

\maketitle
%-------------------> SEC: introduction
\section{Introduction}
\label{sec:intro}

Ultralight bosonic dark matter, also known as Bose--Einstein condensate dark matter (BECDM) or fuzzy dark matter, provides a well--motivated framework in which dark matter is described by a coherent macroscopic wave function governed by the Gross Pitaevskii Poisson (GPP) system of equations. This model has been extensively explored as an alternative to cold dark matter at large and small scales,  whose most important property is that wave effects can suppress small-scale structure and its collapse leads to the formation of solitonic cores \cite{Matos:2000ss,Hu:2000ke,Chavanis2015,Hui:2016,ElisaFerreira,Niemeyer_2020,Hui:2021tkt,10.3389/fspas.2025.1538434}.

Stationary solutions of the GPP system play a central role in this model, as they define important configurations. The simplest of them being the ground-state solution \cite{Ruffini:1969,GuzmanUrena2004}, which acts as an attractor of this dark matter under very general initial conditions, from simple spherical and axisymmetric cases \cite{GuzmanUrena2006,BernalGuzman2006b}, until scenarios of structure formation (e.g. \cite{Schive:2014dra,Mocz:2017wlg,Veltmaat_2018,Gotinga2022}), as well as at local single structure collapse gravitational cooling or kinetic relaxation \cite{Rusos2018,Eggemeier2019,Chen2021} or multi-core mergers \cite{DuNiemeyer2017,Luna2023,periodicas}. Their properties, including scaling relations, stability criteria, and maximum mass configurations, have been widely studied both numerically and analytically \cite{Chavanis_2019,Chavanis_2020,Chen2021,Alvarez_Rios_2023, CarlosIvanFranciscoUniverse}.

Beyond these spherical ground-state solutions there are others that may play an astrophysically relevant role, like the mixed state solutions called gravitational atoms \cite{GuzmanUrena2020}, which have some chances to explain the anisotropy of satellite galaxies around major galaxies \cite{jordi}. However, there are other intriguing solutions that may imprint some observational effects that may support or rule out BECDM, these are the vortex solutions. These states carry quantized angular momentum and exhibit a density depletion along the symmetry axis. Vortex configurations have been studied extensively in laboratory Bose-Einstein condensates \cite{Matthews1999,Leanhardt2002, Fetter2009,PethickSmith}, and have entered the astrophysical context within BECDM, where they may arise in rotating halos or during dynamical processes \cite{Rindler_Daller_2012,Kain_2010,Korshynska_2023,Nikolaieva_2023, Alexander:2021zhx}. Recent studies have also explored their stability and potential observational signatures in galactic systems \cite{Alvarez_Rios_2025,Korshynska_2025,Glennon2023}. However, their structure and stability remain less well understood than those of non-rotating solitonic cores.

On the other hand, an additional ingredient of astrophysical relevance is the presence of central compact objects. Observations indicate that most galaxies host supermassive black holes at their centers \cite{Ghez2008,Gillessen2009,Gravity2019}, and their interaction with ultralight dark matter has been the subject of increasing interest. Previous works have studied accretion processes, dynamical friction, and the impact of black holes on scalar field configurations \cite{Hertzberg2020,Cardoso2022,Boudon2023,Ravanal2023,Wang2022,palomareschavez2025}. In fact, recent numerical studies have investigated the interplay between black holes and BECDM in dynamical scenarios, including mergers and binary systems \cite{Bamber2023,Aurrekoetxea2024,Aurrekoetxea_2024_selfinteracting, Bromley2024}.
A key point in these analyses is the interplay between the black-hole horizon scale and the characteristic length scale of the dark matter, namely the de Broglie wavelength. For ultralight bosons in the mass range relevant to BECDM, this wavelength is typically in the range $\sim 10\,{\rm pc}$ to $\sim 1\,{\rm kpc}$, depending on the boson velocity dispersion and mass. As a consequence, the black-hole horizon is generically much smaller than the coherence scale of the condensate, and local horizon-scale physics is expected to have a negligible impact on the large-scale dynamics of the dark matter wave function.  
In this regime, the black hole can be consistently modeled as an effective point-like Newtonian source, whose effect enters through the external gravitational potential rather than through detailed horizon-scale interactions and accretion. This justifies our treatment of the black hole as a Newtonian point mass throughout this work.

In this framework, we study the interplay between central black holes and vortex solutions of BECDM.  We construct families of stationary axisymmetric solutions of the GPP system in the presence of a central black hole, focusing on both solitonic cores ($m=0$) as a test case and the most relevant scenario involving line-vortex configurations ($m=1$). Using an imaginary-time evolution method, we construct equilibrium solutions across a broad region of parameter space in terms of self-interaction and black hole mass, and analyze their structure and stability.

We make three main contributions. First, we provide a unified description of stationary configurations including both non-rotating and rotating states in the presence of a central potential sourced by a point-like mass. Second, we characterize their stability using scale-invariant parametrizations based on the turning-point criterion \cite{Chavanis:2011,PhysRevD.84.043532}. Finally, we derive empirical scaling relations for the maximum mass and the onset of instability as functions of the black-hole mass.

By showing that there are stable solutions for a vortex of BECDM around a supermassive black hole,  for a wide range of parameters, we motivate the study of potential formation processes and astrophysical observable signatures.

The paper is organized as follows. In Sec. \ref{sec:model}, we introduce the Gross--Pitaevskii--Poisson system and its stationary
equations, including the associated energy functional and scaling properties. In Sec. \ref{sec:numerical-framework}, we describe the
numerical method based on imaginary--time evolution. The main results, including the structure and stability of the stationary solutions, are presented in Sec. \ref{sec:results}. Finally, our conclusions are summarized in Sec. \ref{sec:conclusions}.

%-------------------> SEC: model and equations
\section{Model and equations}
\label{sec:model}

%--------------------------------------------------
\subsection{GPP Evolution Equations and Units}

In our system, Ultralight Bose-Einstein Condensate Dark Matter (BECDM) evolves under its self-gravity and that of the gravitational field due to a point particle that models a central black hole. Thus, the dynamics of the condensate is governed by the coupled Gross--Pitaevskii--Poisson (GPP) system with the potential due to the black hole:

\begin{equation}
\begin{aligned}
i\hbar \frac{\partial \tilde{\Psi}}{\partial \tilde{t}} &=-\frac{\hbar^2}{2m_B}\tilde{\nabla}^2 \tilde{\Psi}
+ m_B \left( \tilde{V} + \tilde{V}_{\rm BH} \right)\tilde{\Psi}+ \tilde{g}\,|\tilde{\Psi}|^2 \tilde{\Psi}, \\
\tilde{\nabla}^2 \tilde{V} &= 4\pi G \tilde{\rho} ,
\end{aligned}
\nonumber %\label{eq:GPP-physical}
\end{equation}

\noindent where $\tilde{\Psi}$ is the macroscopic wave function or order parameter of the condensate, $m_B$ denotes the boson mass, the nonlinear coupling $\tilde{g} = \frac{4\pi\hbar^2 \tilde{a}_s}{m_B^{2}}$ models the short-range self-interactions characterized by the $s$-wave scattering length $\tilde{a}_s$, the gravitational potential $\tilde{V}$ is sourced by the condensate's density $\tilde{\rho} = m_B|\tilde{\Psi}|^2$, while the external potential $\tilde{V}_{\rm BH} = -G\tilde{M}_{\rm BH}/\tilde{r}$ represents the gravitational field of a central black hole of mass $\tilde{M}_{\rm BH}$.

%--------------------------------------------------
%\subsection{Dimensionless Gross--Pitaevskii--Poisson system}

As usual, the GPP system is rewritten in dimensionless code units, for which we introduce a characteristic length scale $x_0=\mathrm{kpc}/\lambda$ and parametrize the boson mass as $m_B=m_{22}\times10^{-22}\,\mathrm{eV}/c^2$. In this way, following \cite{Alvarez_Rios_2025}, we define the rescaled variables

\begin{align}
\tilde{\vec{x}} &= x_0\,\vec{x}, &
\tilde{t} &= t_0\, t, &
\tilde{\rho} &= \frac{M_0}{x_0^3}\, \rho, \nonumber \\
\tilde{V} &= v_0^2\, V, &
\tilde{a}_s &= a_0\, a_s, &
\tilde{M}_{\rm BH} &= M_0\, M_{\rm BH},
\label{eq:scales-def}
\end{align}

\noindent where the associated characteristic scales are

\begin{align}
t_0 &= \frac{m_B x_0^{\,2}}{\hbar}
\simeq 5.096\times10^{-2}
\left(\frac{m_{22}}{\lambda^2}\right)\,\mathrm{Gyr}, \nonumber \\
v_0 &= \frac{\hbar}{m_B x_0}
\simeq 19.20
\left(\frac{m_{22}}{\lambda}\right)
\,\mathrm{km}\,\mathrm{s}^{-1}, \nonumber \\
M_0 &= \frac{\hbar^2}{4\pi G m_B^2 x_0}
\simeq 6.82\times10^{6}
\left(\frac{\lambda}{m_{22}^{2}}\right)\,
\mathrm{M}_\odot, \nonumber \\
a_0 &= \frac{G m_B^{3} x_0^{2}}{\hbar^2}
\simeq 3.228\times10^{-75}
\left(\frac{m_{22}^{3}}{\lambda^2}\right)\,\mathrm{cm}.\nonumber
\end{align}

\noindent Finally, with these definitions, the dimensionless GPP equations become

\begin{equation}
\begin{aligned}
i\,\frac{\partial \Psi}{\partial t} &= H(\Psi)\,\Psi, \\
\nabla^2 V &= \rho ,
\end{aligned}
\label{eq:GPP-dimensionless}
\end{equation}

\noindent where the Hamiltonian operator is given by

\begin{equation}
H(\Psi)= -\frac{1}{2}\nabla^2 + V + V_{\rm BH} + g\,|\Psi|^2,
\nonumber %\label{eq:H-dimensionless}
\end{equation}

\noindent and the black hole potential reads $V_{\rm BH}=-M_{\rm BH}/(4\pi r)$.

In our analysis we also exploit the following scaling symmetry of the GPP system, that includes the black hole mass as follows:

\begin{eqnarray}
%\begin{matrix}
&&    \{t,\vec{x},\Psi,V,g,M_{\rm BH}\}\;\longrightarrow\; \nonumber\\
&&\nonumber\\
&&\{\lambda^{-2}t,\lambda^{-1}\vec{x},
\lambda^{2}\Psi,\lambda^{2}V,
\lambda^{-2}g,\lambda\,M_{\rm BH}\},
\label{eq:scale-symmetry}
%\end{matrix}
\end{eqnarray}

\noindent which leaves the system (\ref{eq:GPP-dimensionless}) invariant.

%--------------------------------------------------
\subsection{Stationary system, energy functional, and stability criteria}

Stationary solutions of the GPP system are obtained by assuming harmonic time dependence of the condensate wave function,

\begin{equation}
\Psi(\vec{x},t)=e^{-i\mu t}\,\psi(\vec{x}),\nonumber
\end{equation}

\noindent where $\mu$ denotes the dimensionless eigenenergy. Substituting this ansatz into Eq.~(\ref{eq:GPP-dimensionless}) leads to the following nonlinear eigenvalue problem

\begin{equation}
H(\psi)\,\psi = \mu\,\psi ,
\label{eq:GPE-stationary-H}
\end{equation}

\noindent constrained by the Poisson equation $\nabla^2 V = |\psi|^2$.

{\it Energy functional.} Equation~(\ref{eq:GPE-stationary-H}) is the result of a constrained variational principle, and stationary solutions correspond to extrema of the enthalpy functional

\begin{equation}
\mathcal{H}[\psi] = E[\psi] - \mu\,N[\psi],\nonumber
\end{equation}

\noindent where $N$ is the particle number,

\begin{equation}
N[\psi]=\int |\psi|^2\,d^3x,\nonumber
\end{equation}

\noindent which in this case coincides with the total mass $M\equiv N$ of the stationary solution. For our system, the total energy has different contributions

\begin{equation}
E[\psi]=K[\psi]+I[\psi]+W[\psi]+W_{\rm BH}[\psi],\nonumber
\end{equation}

\noindent where

\begin{align}
K[\psi] &= \int \frac{1}{2}|\nabla\psi|^2\,d^3x, \nonumber\\
I[\psi] &= \int \frac{g}{2}|\psi|^4\,d^3x, \nonumber\\
W[\psi] &= \int \frac{1}{2}V|\psi|^2\,d^3x, \nonumber\\
W_{\rm BH}[\psi] &= \int V_{\rm BH}|\psi|^2\,d^3x ,\nonumber
\end{align}

\noindent corresponding to the kinetic, self-interaction, self-gravitational, and black hole gravitational energy contributions, respectively. The stationarity condition $\delta\mathcal{H}=0$ recovers the eigenvalue equation (\ref{eq:GPE-stationary-H}). This decomposition is the basis of the numerical method used to construct the solutions below.

{\it Turning points.} The scaling symmetry (\ref{eq:scale-symmetry}) that leaves the GPP system invariant induces the transformation

\begin{equation}
\psi_\lambda(\vec{x})= \lambda^{2}\psi(\lambda^{-1}\vec{x}),
\label{eq:lambda-scaling}
\end{equation}

\noindent which maps stationary configurations into others physically equivalent and the relevant quantities scale as

\begin{eqnarray}
%\begin{matrix}
  &&  \{M,\mu,K,I,W,W_{\rm BH},E\}
~~~\rightarrow \nonumber\\
&&\nonumber\\
&&\{\lambda M,\lambda^{2}\mu,
\lambda^{3}K,\lambda^{3}I,
\lambda^{3}W,\lambda^{3}W_{\rm BH},
\lambda^{3}E\},\label{eq:scale-symmetry2}
%\end{matrix}
\end{eqnarray}

\noindent which implies that the combination $\sqrt{|g|}\,M$ is invariant under rescaling, and therefore provides a scale-independent characterization of families of stationary solutions.
 
Along a continuous family of stationary solutions parametrized by $\mu$, stability can change at turning points, where the mapping between $\mu$ and the conserved quantity $M$ has a maximum value, leading to the Vakhitov-Kolokolov condition \cite{VakhitovKolokolov1973}

\begin{equation}
\frac{dM}{d\mu}=0 .
\label{eq:turning-point}
\end{equation}

\noindent Since $\sqrt{|g|}\,M$ is scale invariant, an equivalent condition can be expressed as

\begin{equation}
\frac{d(\sqrt{|g|}\,M)}{d(g\,\psi_{0})}=0,
\label{eq:maxgM}
\end{equation}

\noindent where $\psi_{0}=\max\{|\psi|\}$. This means that critical points of $\sqrt{|g|}\,M$ as a function of $g\,\psi_{0}$ separate stable and unstable branches of stationary solutions. This turning point criterion has been widely used in the analysis of self-gravitating condensates and boson star configurations  in Refs. \cite{Chavanis:2011,PhysRevD.84.043532} and later on in Refs. \cite{Chen2021,Nikolaieva_2021,Chavez_Nambo_2024,CarlosIvanFranciscoUniverse}. 

%--------------------------------------------------
\subsection{Axisymmetric stationary configurations}

Axisymmetric stationary configurations of the GPP system are characterized by the integer {\it winding number} $m$, which labels the topological charge. For each value of $m$, stationary solutions are solutions of nonlinear eigenvalue problems sharing the same differential operator but subject to different boundary conditions at the axis of symmetry. We denote the eigenfunctions and eigenvalues of such problems by $\phi_m$ and $\mu_m$, respectively.

In cylindrical coordinates $(r_\perp,\varphi,z)$, the stationary wave function can be written as \cite{Korshynska_2023, Korshynska_2025, Glennon:2023oqa, Nikolaieva_2021, Nikolaieva_2023, Alvarez_Rios_2025}:

\begin{equation}
\psi_m(r_\perp,\varphi,z)=\phi_m(r_\perp,z)\,e^{im\varphi},\nonumber
\end{equation}

\noindent where $\phi_m(r_\perp,z)$ is real and axisymmetric. Substitution into the stationary Gross-Pitaevskii equation yields the nonlinear eigenvalue problem

\begin{equation}
H_m(\phi_m)\,\phi_m = \mu_m\,\phi_m ,
\nonumber %\label{eq:axisymmetric-eigenproblem}
\end{equation}

\noindent with the Hamiltonian operator given by

\begin{eqnarray}
H_m(\phi_m) & = & -\frac{1}{2}\left(
\frac{\partial^2}{\partial r_\perp^2}+ \frac{1}{r_\perp}\frac{\partial}{\partial r_\perp} + \frac{\partial^2}{\partial z^2}
- \frac{m^2}{r_\perp^2}\right) \nonumber \\
& +& V + V_{\rm BH} + g\,\phi_m^2 .
\nonumber %\label{eq:Hm-operator}
\end{eqnarray}

\noindent The difference between a \textit{solitonic core} solution and a \textit{line-vortex} solution is determined by the boundary condition at the symmetry axis $r_\perp=0$. For vanishing winding number ($m=0$), regularity allows a finite central density,
$\phi_0(0,0)$, along with the conditions

\begin{eqnarray}
\partial_z\phi(r_\perp,z=0)&=&0,\nonumber\\
\partial_{r_\perp}\phi(r_\perp=0,z) &=&0
%\phi_0(0,z)=\phi_{0},
\label{eq:bcmeq0},
\end{eqnarray}

\noindent at the equatorial plane and the axis respectively.  This condition is sufficient even with the potential $V_{BH}$ as demonstrated in the Appendix.

Now, in the case of nonzero winding number ($m\neq0$), regularity of the wave function in the presence of quantized circulation requires the condition

\begin{equation}
\phi_m(0,z)=0 ,\label{eq:bcmne0}
\end{equation}

\noindent which enforces a zero density along the symmetry axis and defines a \textit{line-vortex}. 

In both cases, localization of the solution requires $\phi_m\to0$ as $r_\perp^2+z^2\to\infty$. This isolation condition allows the gravitational potential $V$ to be determined from the Poisson equation sourced by $\rho=\phi_m^2$. These conditions suffice to guarantee regularity at the origin as shown in the Appendix, also for $m > 0$.

This formulation shows that \textit{solitonic cores} and \textit{line vortices} are solutions of the same stationary GPP system, differing only in their topological charge and the corresponding boundary condition at the symmetry axis. In the following section we describe the numerical framework used to construct these stationary solutions and to analyze their stability.

%-------------------------------------> SEC: Numerical framework
\section{Numerical framework}
\label{sec:numerical-framework}

\subsection{Numerical methods}

{\it Imaginary-time evolution method.} Stationary axisymmetric configurations of the GPP system are constructed using an imaginary-time evolution method \cite{Alvarez_Rios_2025}, which relaxes the system toward stationary solutions by minimizing the energy functional at fixed normalization. All configurations, including \textit{solitonic cores} ($m=0$) and \textit{line vortices} ($m\neq0$), are obtained by evolving the same stationary equation, just using the two mentioned boundary conditions (\ref{eq:bcmeq0}) and (\ref{eq:bcmne0}) respectively.

Working in dimensionless units and defining the imaginary time $\tau = i t$, it is possible to define an evolution equation for the stationary amplitude $\phi_m$:

\begin{equation}
\frac{\partial \phi_m}{\partial \tau}= -\left[ H_m(\phi_m) - \mu_m \right]\phi_m ,
\label{eq:imaginary-GPE}
\end{equation}

\noindent where $H_m$ is the nonlinear Hamiltonian operator associated with winding number $m$. Equation~(\ref{eq:imaginary-GPE}) is discretized in space using second-order finite differences and integrated in imaginary time using a first-order explicit Euler scheme,

\begin{equation}
\phi_m^{n+1} = \phi_m^{n} - \Delta\tau\, \left[ H_m(\phi_m^{n}) - \mu_m \right]\phi_m^{n}.
\nonumber
\end{equation}

\noindent This evolution can be interpreted as a gradient-descent method in function space used to solve the stationary problem. In analogy with finding a root of a function via gradient methods, in our case the goal is to find a configuration $\phi_m$ such that

\begin{equation}
\frac{\delta \mathcal{H}}{\delta \phi_m} = 0,\nonumber
\end{equation}

\noindent where $\mathcal{H}=E-\mu N$ is the constrained functional to be minimized. The functional derivative reads

\begin{equation}
\frac{\delta \mathcal{H}}{\delta \phi_m} = \frac{\delta E}{\delta \phi_m}-\mu\,\frac{\delta N}{\delta \phi_m} = 
H_m(\phi_m)\phi_m - \mu_m \phi_m .\nonumber
\end{equation}

\noindent and therefore, Eq.~(\ref{eq:imaginary-GPE}) can be written as

\begin{equation}
\frac{\partial \phi_m}{\partial \tau} =  - \frac{\delta \mathcal{H}}{\delta \phi_m},\nonumber
\end{equation}

\noindent which shows that the imaginary-time evolution drives the system toward a critical point of $\mathcal{H}$, i.e., a solution of the nonlinear eigenvalue problem $H_m(\phi_m)\phi_m=\mu_m\phi_m$.

At each imaginary--time step the wave function is explicitly rescaled to fix its maximum amplitude,

\begin{equation}
\phi_m \rightarrow \frac{\phi_m}{\max(\phi_m)} .\nonumber
\end{equation}

This procedure exploits the scaling symmetry of the GPP system (\ref{eq:scale-symmetry}) and allows all stationary solutions to be constructed with $\phi_0 :=\psi_{0}=1$. Physical solutions with arbitrary amplitudes are then recovered after each time-step by applying the corresponding scaling transformation.

The imaginary-time evolution is continued until convergence is achieved, as determined by the residual of the stationary equation,

\begin{equation}
\| H_m(\phi_m)-\mu_m \phi_m \|_2 < 10^{-4}.\nonumber
\end{equation}

\noindent\textit{Fourier--space Poisson solver.} 
At each iteration during the evolution, we obtain the gravitational potential $V$ in the Hamiltonian $H_m$ by solving the Poisson equation:

\begin{equation}
\left(
\frac{\partial^2}{\partial r_\perp^2}
+\frac{1}{r_\perp}\frac{\partial}{\partial r_\perp}
+\frac{\partial^2}{\partial z^2}
\right)V = \rho, ~~~ \rho = \phi_m^2.
\label{eq:poisson-axisymmetric}
\end{equation}

\noindent for which we apply a Fourier transform along the $z-$direction,

\begin{equation}
\hat{V}(r_\perp,k_z) =\mathcal{F}_z[V] = \int V(r_\perp,z)e^{ik_z z}\,dz,\nonumber
\end{equation}

\noindent and similarly $\hat{\rho}=\mathcal{F}_z[\rho]$. Specifically we compute the transform and its inverse using the fast Fourier transform (FFT) as illustrated in \cite{periodicas, rodriguezlara2024}. In Fourier space, Eq.~(\ref{eq:poisson-axisymmetric}) reduces to

\begin{equation}
\frac{d^2\hat{V}}{dr_\perp^2}  +\frac{1}{r_\perp}\frac{d\hat{V}}{dr_\perp} -k_z^2\hat{V} = \hat{\rho}.
\nonumber %\label{eq:poisson-fourier}
\end{equation}

\noindent that we solve on the discrete radial domain given by $r_i=i\,\Delta r_\perp$, which yields a tridiagonal system:

\begin{equation}
a_i \hat{V}_{i-1}+b_i \hat{V}_i+c_i \hat{V}_{i+1} = \hat{\rho}_{i},\nonumber
\end{equation}

\noindent with $a_i=\frac{1}{\Delta r_\perp^2}-\frac{1}{2r_i\Delta r_\perp}$, $b_i=-\frac{2}{\Delta r_\perp^2}-k_z^2$, and $c_i=\frac{1}{\Delta r_\perp^2}+\frac{1}{2r_i\Delta r_\perp}$. For each Fourier mode, the system is solved using the Thomas algorithm \cite{Thomas2}, and the solution $V(r_\perp,z)$ is reconstructed via the inverse FFT.

Boundary conditions are imposed as follows. Regularity at the axis requires

\begin{equation}
\left.\frac{d\hat{V}}{dr_\perp}\right|_{r_\perp=0}=0,\nonumber
\end{equation}

\noindent while at $r_\perp=r_{\max}$ we impose a monopolar boundary condition

\begin{equation}
V(r_\perp,z)\big|_{r_\perp=r_{\max}} = -\frac{M}{4\pi\sqrt{r_\perp^2+z^2}}.\nonumber
\end{equation}

\noindent which have to be fulfilled.

%--------------------------------------------------
\subsection{Numerical setup and parameter space}

All solutions are constructed on a finite two-dimensional domain $(r_\perp,z)\in[0,20]\times[-20,20]$ in dimensional units, uniformly discretized with a $128\times128$ grid and spatial resolution $\Delta z = 2\Delta r_\perp = 0.3125$. The imaginary-time step is chosen according to a parabilic stability condition

\begin{equation}
\Delta\tau = 0.25\,\min(\Delta r_\perp,\Delta z)^2 = 2.5\times10^{-2},\nonumber
\end{equation}

\noindent which ensures stability and convergence of the explicit Euler scheme.

{\it Parameter space.} We construct stationary configurations only for winding numbers $m=0$ and $m=1$, corresponding to solitonic core and singly quantized line-vortex configurations. The self-interaction coefficient is varied in the range

\begin{equation}
g\in[-2,-0.5),
\nonumber %\label{eq:g-range}
\end{equation}

\noindent while the dimensionless black-hole mass covers the range

\begin{equation}
M_{\rm BH}\in[0,20).
\nonumber %\label{eq:MBH-range}
\end{equation}

\noindent In order to provide a physical reference, we recall that the dimensionless quantities are related to their physical counterparts through the scaling relations (\ref{eq:scales-def}). For the fiducial values $\lambda=1$ and $m_{22}=1$, the characteristic mass scale is $M_0 \simeq 6.82\times10^{6}\,M_\odot$. The explored interval $M_{\rm BH}\in[0,20]$ therefore corresponds to physical black-hole masses in the range

\begin{equation}
\tilde{M}_{\rm BH}\in[0,1.36\times10^{8}]\,{\rm M}_\odot ,\nonumber
\end{equation}

\noindent which goes from the absence of a central black hole up to masses typical of supermassive black holes in galactic nuclei. In particular, values $M_{\rm BH}\sim 1$ correspond to $\tilde{M}_{\rm BH}\sim 7\times10^{6}\,M_\odot$, comparable to the mass of the black hole at the center of the Milky Way ($\mathrm{Sgr\,A^\ast}$) \cite{Ghez2008,Gillessen2009,Gravity2019}, while the upper end of the interval reaches the regime of intermediate and moderately massive supermassive black holes.

Another boundary of our parameter space is the restriction to ground states. Although excited stationary states can, in principle, be obtained within the imaginary-time approach by enforcing orthogonality with respect to previously computed solutions, we restrict our analysis to the ground state for the two cases $m=0,1$ \cite{Chavez_Nambo_2024, Nikolaieva_2023}. This choice is motivated by both numerical and physical considerations. It is well known that ground-state configurations for $m=0$ are dynamical attractors in BECDM simulations \cite{Schive:2014dra,Mocz:2017wlg,Veltmaat_2018,Gotinga2022}, as well as in local collapse simulations \cite{Rusos2018,Chen2021} and not only for pure BECDM, but also in the presence of gas acting as baryonic matter \cite{AlvarezGuzmanNiemeyer_2025}, and in the presence of black holes, which promote condensation toward stationary core solutions in the fuzzy dark matter regime \cite{palomareschavez2025}. In contrast, excited states typically exhibit multiple stability branches \cite{GuzmanUrena2006}, leading to a more intricate stability structure (e. g. \cite{Chavez_Nambo_2024}) whose possible formation -and therefore astrophysical relevance- has not yet been studied.

Finally, we restrict our analysis to the attractive self-interaction regime, $g<0$, which is also motivated by stability considerations. Stationary solutions correspond to critical points of $\mathcal{H}$, with stability determined by the sign of $\delta^2 \mathcal{H}$. A practical criterion is given by the Vakhitov-Kolokolov condition (\ref{eq:turning-point}),$\frac{dN}{d\mu} < 0$, which characterizes stable configurations, whereas $dN/d\mu > 0$ signals instability. While the repulsive case $g>0$ typically satisfies this condition along the ground-state branch, the attractive regime $g<0$ admits unstable configurations, leading to a richer dynamical structure.

Once we have defined the numerical framework and parameter space, we proceed to present the corresponding stationary solutions. The following section analyzes their structural properties and stability as functions of the winding number, self-interaction strength, and black-hole mass.

%--------------------------------------> RESULTS
\section{Results}
\label{sec:results}

\subsection{Structure of stationary solutions}

\begin{figure}
\centering
\includegraphics[width=0.95\linewidth]{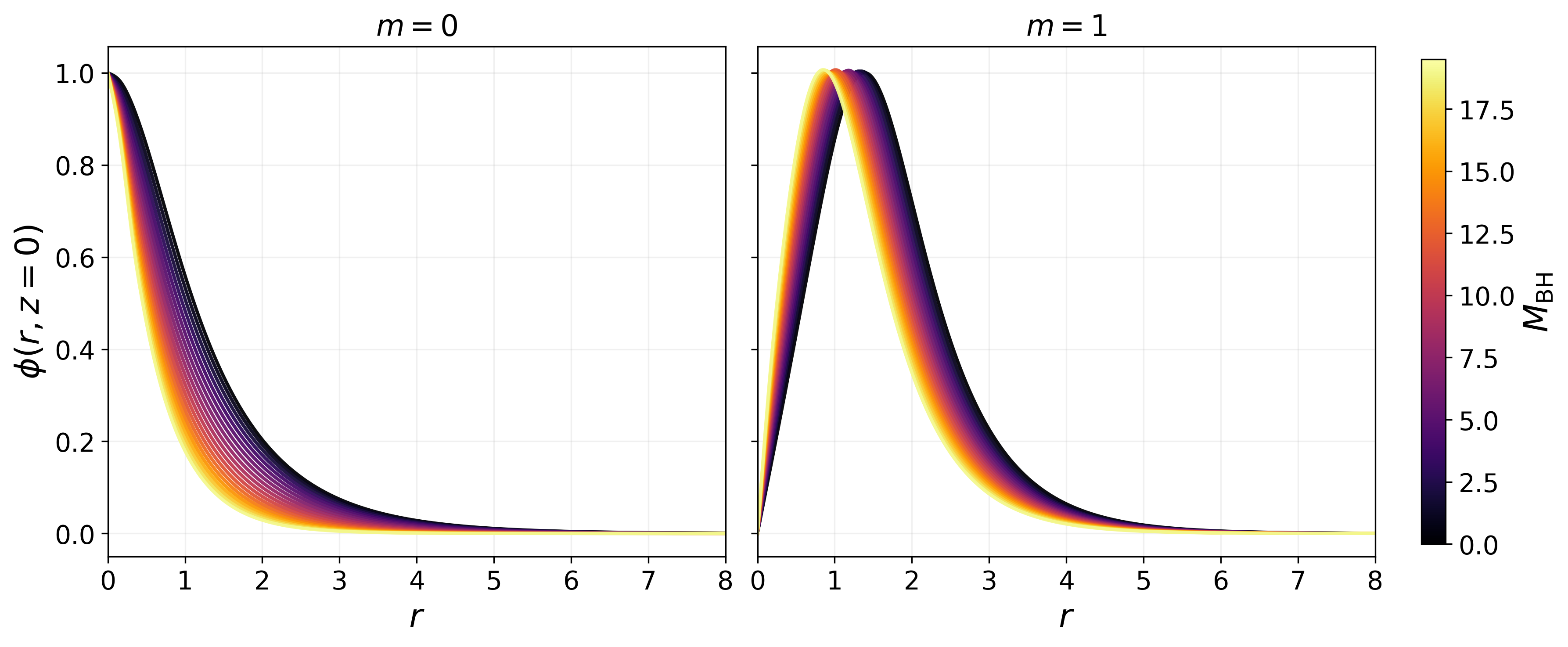}
\caption{Radial profiles on the equatorial plane ($z=0$) for fixed self--interaction $g=-2$ and varying black-hole masses $M_{\rm BH}\in[0,20)$. At the left / right we show the case $m=0$ / $m=1$.}
\label{fig:profiles_MBH}
\end{figure}

\begin{figure}
\centering
\includegraphics[width=0.95\linewidth]{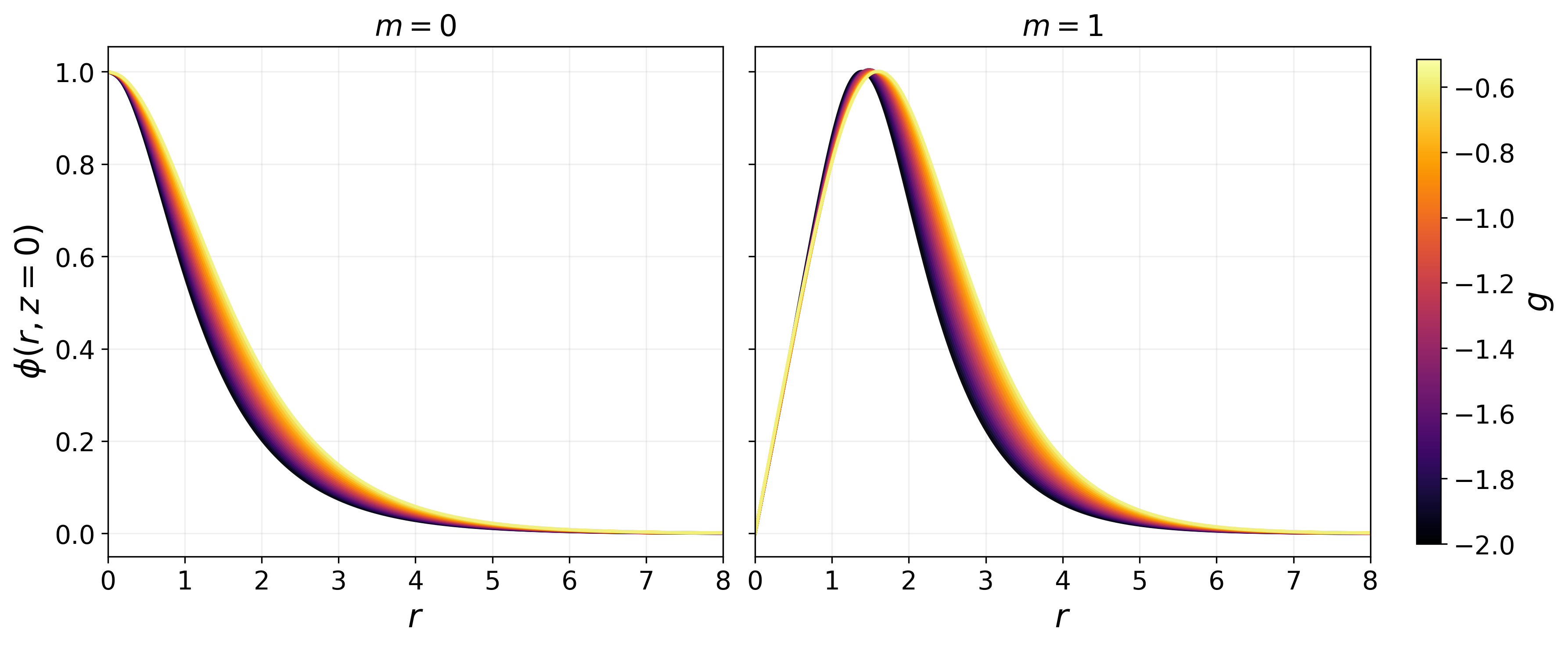}
\caption{Radial profiles at the equatorial plane ($z=0$) with $M_{\rm BH}=0$ for different self-interaction strength values $g\in[-2,0)$. At the left / right we show the case $m=0$ / $m=1$.}
\label{fig:profiles_g}
\end{figure}

We begin by illustrating the structure of the stationary solutions on the equatorial plane ($z=0$). Figures \ref{fig:profiles_MBH} and \ref{fig:profiles_g} show the density profiles along the radial coordinate for both solitonic core ($m=0$) and line-vortex ($m=1$) configurations.

{\it Effect of the black hole mass.} Figure \ref{fig:profiles_MBH} shows solutions  for the self-interaction strength fixed to $g=-2$ and increasing black-hole mass $M_{\rm BH}\in[0,20)$. As $M_{\rm BH}$ increases, the gravitational potential deepens, leading to a systematic contraction of the density profiles. For $m=0$, this produces a stronger central concentration, whereas for $m=1$ the density peak shifts inward while preserving the characteristic zero density at the symmetry axis.

{\it Effect of self-interaction.} Figure \ref{fig:profiles_g} shows solutions  in the absence of a central black hole ($M_{\rm BH}=0$) for different values of the self-interaction strength $g\in[-2,-0.5)$. As $|g|$ increases, the attractive interaction becomes stronger and the profiles contract progressively. For $m=0$, the configurations become more centrally peaked, while for $m=1$ the density maximum moves toward the axis, maintaining the vortex line at the origin.

In all cases, the qualitative behavior among the cases $m=0$ is the same, namely a finite central density with different compactness, whereas for $m=1$ all solutions show a ring-like distribution of matter that vanishes at the axis due to the vortex nature of the solution.

%--------------------------------------------------

%--------------------------------------------------
%--------------------------------------------------

\subsection{Stability and invariant parametrization}

In order to analyze the stability of the solutions, we exploit the scaling symmetry of the GPP system and construct scale-invariant quantities. In particular, we consider the invariant combinations $\sqrt{|g|}\,M$ and $g\,\phi_{0}$, where $\phi_{0}=\max\{\phi\}$.
Notice that this is only a normalization convention and does not reduce generality. Because the Gross-Pitaevskii-Poisson system has the scaling symmetry (\ref{eq:scale-symmetry}), physical solutions with arbitrary central amplitude can be reconstructed by applying the corresponding rescaling. For this reason, results are reported using the scale-invariant combination $g ~\phi_0$, which is independent of the parameter $\lambda$ in Eq. (\ref{eq:scale-symmetry}), and parametrizes the corresponding scale-equivalent families of stationary solutions generated from our parameter space. Therefore our results can be mapped into the corresponding scale-equivalent families of solutions with arbitrary central field values by using any $\lambda \ne 1$, whose physical properties are determined by the scaling relations (\ref{eq:scale-symmetry}) and (\ref{eq:scale-symmetry2}).

\begin{figure}
\centering
\includegraphics[width=0.9\linewidth]{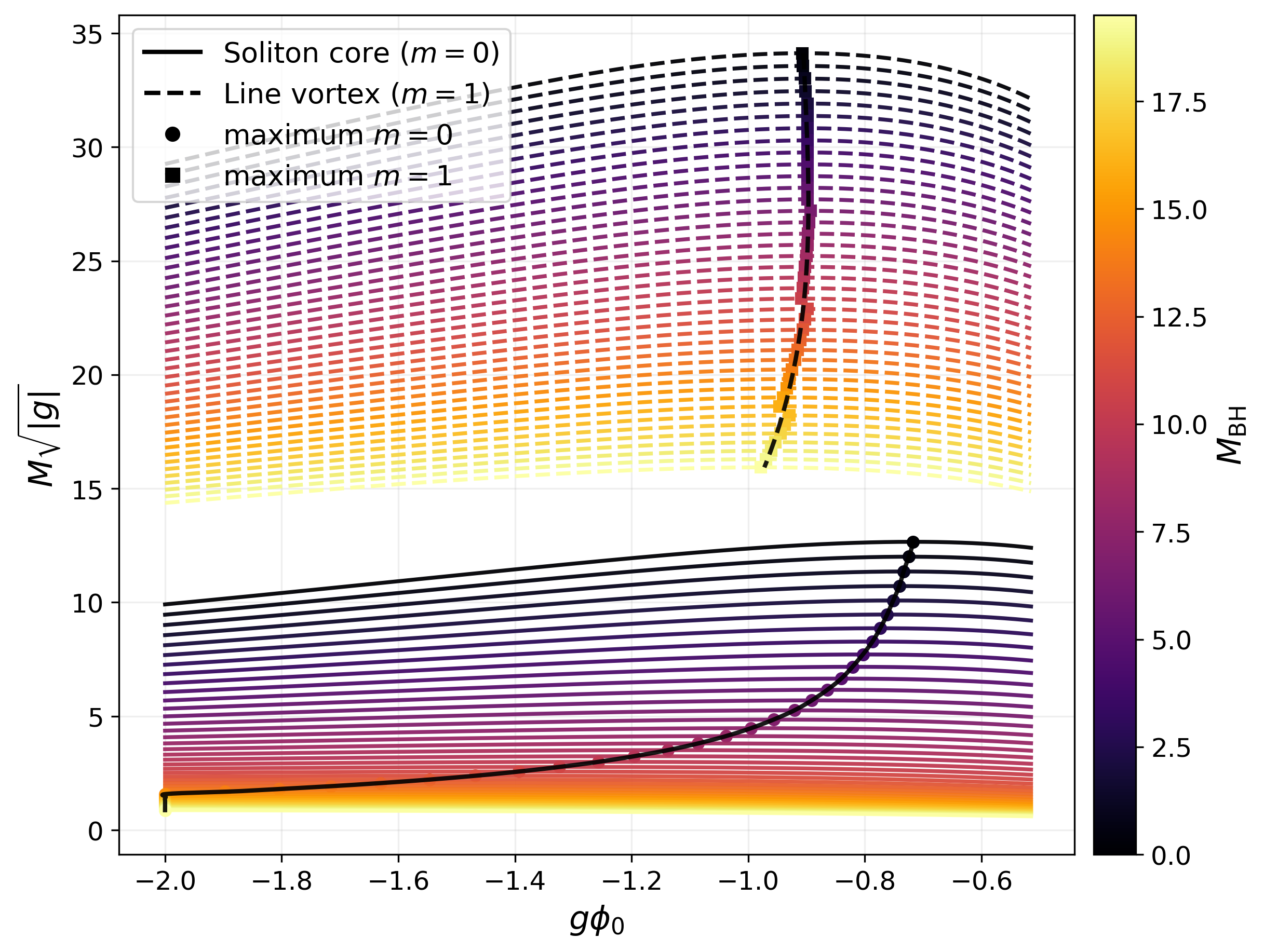}
\caption{Invariant mass $\sqrt{|g|}\,M$ as a function of the invariant central field $g\,\phi_{0}$ for $m=0$ (solid) and $m=1$ (dashed), and different values of the black--hole mass $M_{\rm BH}$ (color scale). The dots indicate the location of the maximum mass along each family of solutions.}
\label{fig:stability}
\end{figure}

In Figure \ref{fig:stability} we show the invariant mass $\sqrt{|g|}\,M$ as a function of $g\,\phi_{0}$ for a number of families corresponding to the solitonic core ($m=0$) and line-vortex ($m=1$) configurations, for different values of the black-hole masses $M_{\rm BH}$.

Notice that for each value of $M_{\rm BH}$, the solutions define a continuous family of solutions in the invariant plane. Each family exhibits a turning point, characterized by an extremum of $\sqrt{|g|}\,M$, where the condition in Eq.~(\ref{eq:maxgM}) is satisfied. According to the turning-point criterion, this critical point separates stable from unstable branches of the family.

For the case $m=0$, the turning point separates a stable branch at lower values of $g\,\phi_{0}$ from an unstable branch at higher values. A similar structure is found for $m=1$, although the corresponding branches are shifted due to the non-zero angular momentum of the solution and the associated vortex structure.

As $M_{\rm BH}$ increases, the branches are systematically displaced in the invariant plane, indicating that the central black hole modifies both the maximum mass and the location of the stability threshold.

%--------------------------------------------------
%--------------------------------------------------

\subsection{Critical points and scaling relations}

The turning points identified in Fig.~\ref{fig:stability} define the maximum--mass configurations along each family of solutions. These points satisfy the condition in Eq.~(\ref{eq:maxgM}) and determine the onset of instability.

\begin{figure}
\centering
\includegraphics[width=0.9\linewidth]{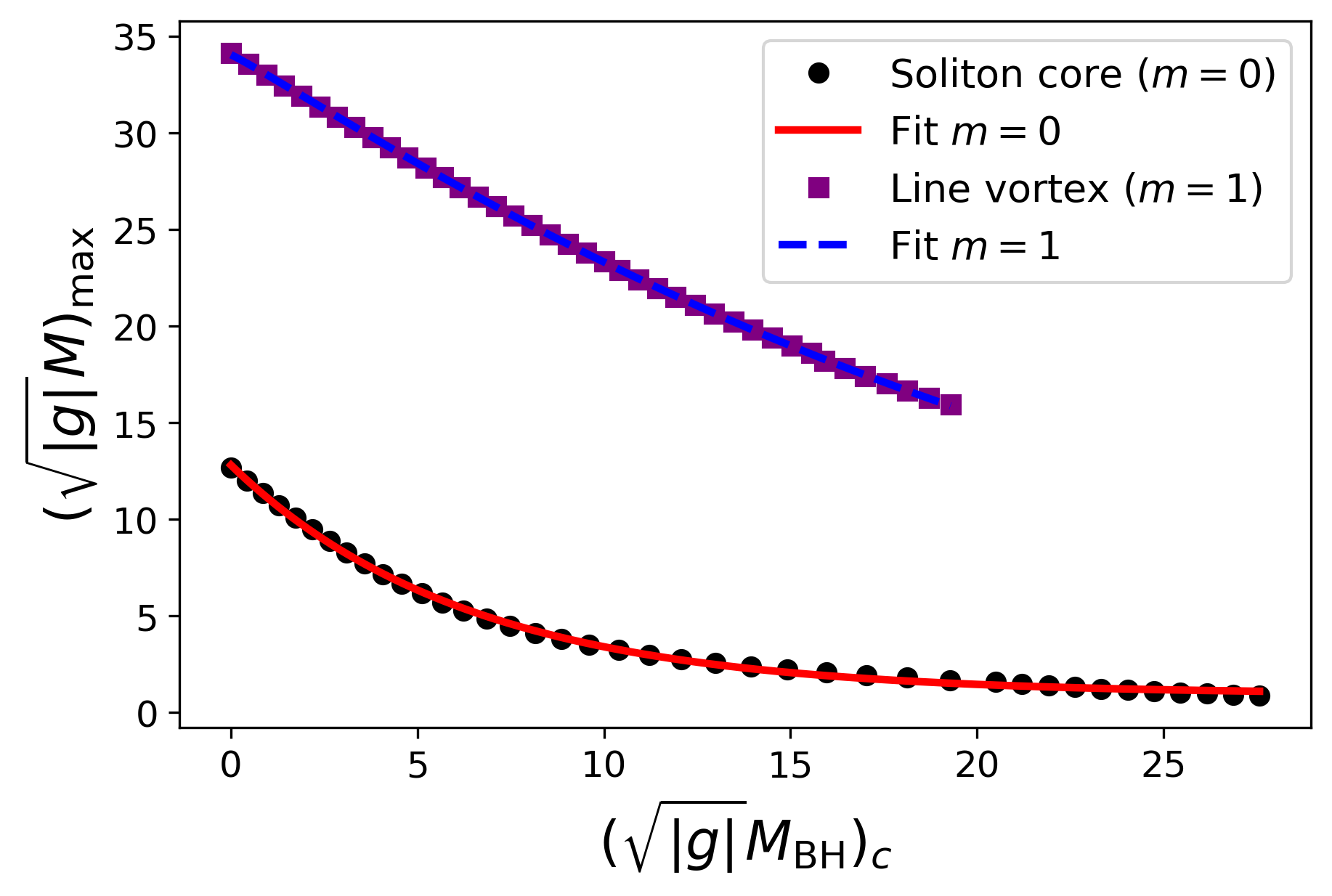}
\caption{Maximum invariant mass $(\sqrt{|g|}M)_{\max}$ as a function of $(\sqrt{|g|}M_{\rm BH})_c$ for $m=0$ (circles) and $m=1$ (squares). Lines show the corresponding fits.}
\label{fig:Mmax_fit}
\end{figure}

\begin{figure}
\centering
\includegraphics[width=0.9\linewidth]{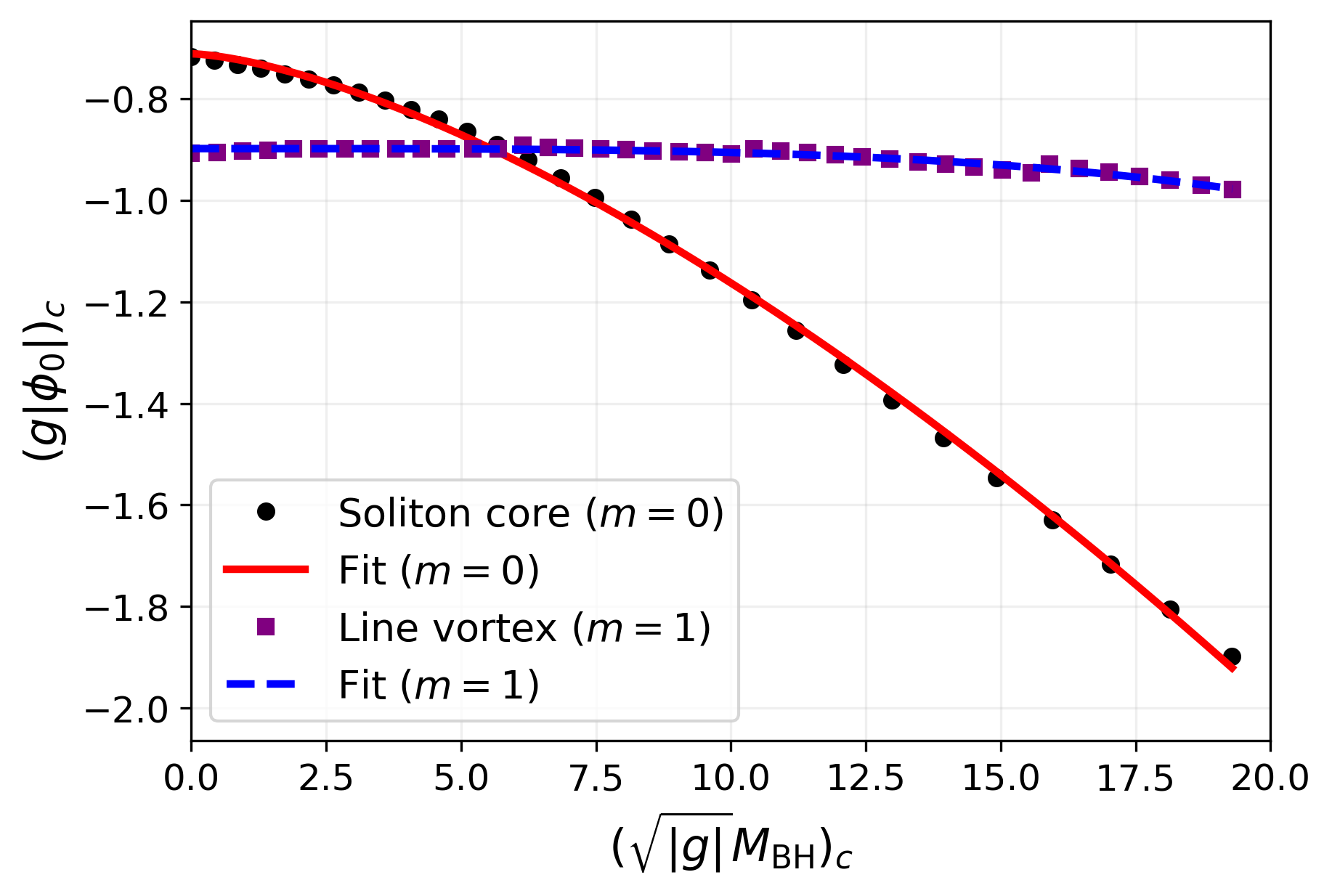}
\caption{Critical invariant $(g\,\phi_{0})_c$ as a function of $(\sqrt{|g|}M_{\rm BH})_c$ for $m=0$ and $m=1$. Lines indicate the best--fit models.}
\label{fig:psimax_fit}
\end{figure}

Extracting these maxima, we construct the dependence of the invariant quantities $(\sqrt{|g|}M)_{\max}$ and $(g\,\phi_{0})_c$ on the critical parameter $(\sqrt{|g|}M_{\rm BH})_c$. The resulting relations are shown in Figs.~\ref{fig:Mmax_fit} and \ref{fig:psimax_fit}.

\noindent\textit{Maximum mass.} Figure~\ref{fig:Mmax_fit} shows $(\sqrt{|g|}M)_{\max}$ as a function of $(\sqrt{|g|}M_{\rm BH})_c$ for both $m=0$ and $m=1$. The numerical data are well described by

\begin{align}
(\sqrt{|g|}M)_{\max}^{(m=0)} &=
12.8 - 11.9\left(1-e^{-0.158\,x^{0.998}}\right), \label{eq:b1}\\
(\sqrt{|g|}M)_{\max}^{(m=1)} &=
34.0 - 33.7\left(1-e^{-0.0325\,x^{1.07}}\right),\label{eq:b2}
\end{align}

\noindent where $x=(\sqrt{|g|}M_{\rm BH})_c$. In both cases, the maximum mass decreases as the black--hole mass increases, indicating that the central potential reduces the amount of mass that can be supported in equilibrium.

However, the relative variation differs significantly between the two configurations. Over the explored range, the normalized change satisfies

\begin{equation}
\frac{\Delta(\sqrt{|g|}M)_{\max}}{(\sqrt{|g|}M)_{\max}} \approx\left\{  
\begin{array}{cc}
0.93, & m=0\\
&\\
0.53, & m=1
\end{array}
\right.
\end{equation}

\noindent showing that solitonic cores experience a bigger relative change in maximum mass across the parameter space.
This indicates that $m=0$ configurations are significantly more sensitive to the central gravitational field, whereas vortex solutions ($m=1$) respond more moderately, consistent with a greater structural robustness likely associated with angular momentum support.

{\it Critical amplitude.} Figure~\ref{fig:psimax_fit} shows the critical invariant $(g\,\phi_{0})_c$ as a function of $(\sqrt{|g|}M_{\rm BH})_c$. The data can be accurately fitted by the formulas

\begin{eqnarray}
(g\,\phi_{0})_c^{(m=0)} &=& -0.712 - 0.0142\,y^{1.50}, \label{eq:pheno1}\\
(g\,\phi_{0})_c^{(m=1)} &=& -0.898 - 2.16\times10^{-6}\,y^{3.55},\label{eq:pheno2}
\end{eqnarray}

\noindent where $y=(\sqrt{|g|}M_{\rm BH})_c$. For $m=0$, the critical amplitude decreases with increasing black-hole mass, indicating that the onset of instability occurs at progressively lower central densities. For $m=1$, the variation is much weaker, reflecting the stabilizing role of angular momentum in vortex configurations. A comment is in order regarding the phenomenological formulas (\ref{eq:b1})–(\ref{eq:pheno2}). These expressions are intended only as empirical fits that describe the behavior of the solutions within the parameter space explored in this work. They are not meant to be used as extrapolations beyond the range of parameters considered in our simulations.

Overall, these results show that the presence of a central black hole modifies both the maximum mass and the location of the stability threshold, with a  stronger effect on solitonic core solutions than on line vortices.

These results are consistent with previously reported limits in specific regimes of the parameter space. In the absence of a central black hole ($M_{\rm BH}=0$), the maximum mass for $m=0$ agrees with our previous determination based on a Sturm-Liouville formulation solved using genetic algorithms \cite{CarlosIvanFranciscoUniverse}, which is in turn consistent with the results obtained via the shooting method in \cite{Chen2021}.

In the complementary limit of vanishing self-interaction ($g=0$) and finite black-hole mass, the resulting profiles are consistent with the stationary solutions reported in Ref.~\cite{palomareschavez2025}, also constructed using shooting techniques.

These consistency regimes validate the imaginary-time approach  used across different regions of the parameter space and confirm its ability to recover known limiting solutions. Moreover, these results are also in agreement with analytical resuilts in \cite{Chavanis_2019,Chavanis_2020}, where an approximate expression of the maximum mass relation is obtained by taking into
account both quantum effects, attractive self-interaction, and self-gravity.

%--------------------------------------------------

%--------------------------------------------------

%--------------------------------------------------
\section{Conclusions}
\label{sec:conclusions}

We have constructed stationary solutions of self-gravitating Bose-Einstein condensate dark matter in the presence of a central black hole, within the Gross-Pitaevskii-Poisson framework for solitonic cores and line-vortex configurations. For this we implemented a novel method that uses imaginary time evolution. We also analyzed stability of the families in a wide parameter space in a scale-invariant formulation.

We have shown that both the black--hole potential and the attractive self--interaction produce a systematic compactification of the solutions. Increasing either $M_{\rm BH}$ or $|g|$ leads to configurations that are more centrally concentrated. However, the physical origin of this behavior differs in each case: the former deepens the external gravitational potential, whereas the latter strengthens the intrinsic self-binding of the condensate.

The stability analysis, based on scale-invariant quantities and the turning-point criterion, reveals that the maximum mass along each family of solutions is strongly affected by the presence of the central black hole. In particular, the invariant mass $(\sqrt{|g|}M)_{\max}$ decreases monotonically with increasing $(\sqrt{|g|}M_{\rm BH})_c$.

Another interesting result is the distinct response of the two cases analyzed. While vortex configurations ($m=1$) exhibit a larger absolute variation in their maximum mass, solitonic cores ($m=0$) experience a stronger relative reduction, indicating a higher sensitivity to the central gravitational field. This behavior suggests that angular momentum provides a mechanism of structural support, making vortex solutions more robust against the influence of the black hole.

We have also shown that the dependence of the maximum mass and the critical amplitude on the black--hole mass can be modeled  by simple empirical relations in terms of scale-invariant variables. These relations provide a compact characterization of the stability properties of the system and may serve as a tool for connecting theoretical models of ultralight dark matter with astrophysical environments hosting central black holes.

Finally a comment on the astrophysical relevance of these configurations is in turn. It has already been shown that solitonic solutions around a black hole can be formed and in fact are attractors within the Fuzzy Dark Matter model under very general conditions \cite{palomareschavez2025}, thereby its relevance in SMBH-FDM astrophysics (e.g \cite{Moczfdmbh,Lancaster_2020,ElZant2020,Hertzberg2020,Wang2022,Alonso-Alvarez2024}). The line-vortex solutions on the other hand, have not been studied in detail in the presence of black holes, and since we have shown now in this paper that they have a stable branch in each family of parameters, there is enough motivation to follow two natural steps including a demonstration of their stability in a full 3D environment and the search of a possible formation mechanism of these solutions, at least locally near the $z=0$ plane.

\section{Data availability}

The numerical results of all the equilibrium configurations constructed in this paper, as well as the data needed to reproduce all the plots are publicly available at \cite{ourdata}.

% ----->     ACKNOWLEDGMENTS     <-----
\section*{Acknowledgments}
This research is supported by  
SECIHTI Grant No. CFB-2025-I-759, 
Laboratorio Nacional de C\'omputo de Alto Desempe\~no Grant No. 2026-8, and 
CIC-UMSNH Grant No. 4.9.

% -------------------------------------------------------
% -----     REFERENCES     ----------
% -------------------------------------------------------
\FloatBarrier
\bibliography{BECDM}

\begin{thebibliography}{67}%
\makeatletter
\providecommand \@ifxundefined [1]{%
 \@ifx{#1\undefined}
}%
\providecommand \@ifnum [1]{%
 \ifnum #1\expandafter \@firstoftwo
 \else \expandafter \@secondoftwo
 \fi
}%
\providecommand \@ifx [1]{%
 \ifx #1\expandafter \@firstoftwo
 \else \expandafter \@secondoftwo
 \fi
}%
\providecommand \natexlab [1]{#1}%
\providecommand \enquote  [1]{``#1''}%
\providecommand \bibnamefont  [1]{#1}%
\providecommand \bibfnamefont [1]{#1}%
\providecommand \citenamefont [1]{#1}%
\providecommand \href@noop [0]{\@secondoftwo}%
\providecommand \href [0]{\begingroup \@sanitize@url \@href}%
\providecommand \@href[1]{\@@startlink{#1}\@@href}%
\providecommand \@@href[1]{\endgroup#1\@@endlink}%
\providecommand \@sanitize@url [0]{\catcode `\\12\catcode `\$12\catcode
  `\&12\catcode `\#12\catcode `\^12\catcode `\_12\catcode `\%12\relax}%
\providecommand \@@startlink[1]{}%
\providecommand \@@endlink[0]{}%
\providecommand \url  [0]{\begingroup\@sanitize@url \@url }%
\providecommand \@url [1]{\endgroup\@href {#1}{\urlprefix }}%
\providecommand \urlprefix  [0]{URL }%
\providecommand \Eprint [0]{\href }%
\providecommand \doibase [0]{http://dx.doi.org/}%
\providecommand \selectlanguage [0]{\@gobble}%
\providecommand \bibinfo  [0]{\@secondoftwo}%
\providecommand \bibfield  [0]{\@secondoftwo}%
\providecommand \translation [1]{[#1]}%
\providecommand \BibitemOpen [0]{}%
\providecommand \bibitemStop [0]{}%
\providecommand \bibitemNoStop [0]{.\EOS\space}%
\providecommand \EOS [0]{\spacefactor3000\relax}%
\providecommand \BibitemShut  [1]{\csname bibitem#1\endcsname}%
\let\auto@bib@innerbib\@empty
%</preamble>
\bibitem [{\citenamefont {Matos}\ and\ \citenamefont
  {Urena-Lopez}(2001)}]{Matos:2000ss}%
  \BibitemOpen
  \bibfield  {author} {\bibinfo {author} {\bibfnamefont {Tonatiuh}\
  \bibnamefont {Matos}}\ and\ \bibinfo {author} {\bibfnamefont {L.~Arturo}\
  \bibnamefont {Urena-Lopez}},\ }\bibfield  {title} {\enquote {\bibinfo {title}
  {{A Further analysis of a cosmological model of quintessence and scalar dark
  matter}},}\ }\href {\doibase 10.1103/PhysRevD.63.063506} {\bibfield
  {journal} {\bibinfo  {journal} {Phys. Rev. D}\ }\textbf {\bibinfo {volume}
  {63}},\ \bibinfo {pages} {063506} (\bibinfo {year} {2001})}\BibitemShut
  {NoStop}%
%%CITATION = ASTRO-PH/0006024;%%
\bibitem [{\citenamefont {Hu}\ \emph {et~al.}(2000)\citenamefont {Hu},
  \citenamefont {Barkana},\ and\ \citenamefont {Gruzinov}}]{Hu:2000ke}%
  \BibitemOpen
  \bibfield  {author} {\bibinfo {author} {\bibfnamefont {Wayne}\ \bibnamefont
  {Hu}}, \bibinfo {author} {\bibfnamefont {Rennan}\ \bibnamefont {Barkana}}, \
  and\ \bibinfo {author} {\bibfnamefont {Andrei}\ \bibnamefont {Gruzinov}},\
  }\bibfield  {title} {\enquote {\bibinfo {title} {{Cold and fuzzy dark
  matter}},}\ }\href {\doibase 10.1103/PhysRevLett.85.1158} {\bibfield
  {journal} {\bibinfo  {journal} {Phys. Rev. Lett.}\ }\textbf {\bibinfo
  {volume} {85}},\ \bibinfo {pages} {1158--1161} (\bibinfo {year}
  {2000})}\BibitemShut {NoStop}%
%%CITATION = ASTRO-PH/0003365;%%
\bibitem [{\citenamefont {Chavanis}(2015)}]{Chavanis2015}%
  \BibitemOpen
  \bibfield  {author} {\bibinfo {author} {\bibfnamefont {Pierre-Henri}\
  \bibnamefont {Chavanis}},\ }\enquote {\bibinfo {title} {Self-gravitating
  bose-einstein condensates},}\ in\ \href {\doibase
  10.1007/978-3-319-10852-0_6} {\emph {\bibinfo {booktitle} {Quantum Aspects of
  Black Holes}}},\ \bibinfo {editor} {edited by\ \bibinfo {editor}
  {\bibfnamefont {Xavier}\ \bibnamefont {Calmet}}}\ (\bibinfo  {publisher}
  {Springer International Publishing},\ \bibinfo {address} {Cham},\ \bibinfo
  {year} {2015})\ pp.\ \bibinfo {pages} {151--194}\BibitemShut {NoStop}%
\bibitem [{\citenamefont {Hui}\ \emph {et~al.}(2017)\citenamefont {Hui},
  \citenamefont {Ostriker}, \citenamefont {Tremaine},\ and\ \citenamefont
  {Witten}}]{Hui:2016}%
  \BibitemOpen
  \bibfield  {author} {\bibinfo {author} {\bibfnamefont {Lam}\ \bibnamefont
  {Hui}}, \bibinfo {author} {\bibfnamefont {Jeremiah~P.}\ \bibnamefont
  {Ostriker}}, \bibinfo {author} {\bibfnamefont {Scott}\ \bibnamefont
  {Tremaine}}, \ and\ \bibinfo {author} {\bibfnamefont {Edward}\ \bibnamefont
  {Witten}},\ }\bibfield  {title} {\enquote {\bibinfo {title} {Ultralight
  scalars as cosmological dark matter},}\ }\href {\doibase
  10.1103/PhysRevD.95.043541} {\bibfield  {journal} {\bibinfo  {journal} {Phys.
  Rev. D}\ }\textbf {\bibinfo {volume} {95}},\ \bibinfo {pages} {043541}
  (\bibinfo {year} {2017})}\BibitemShut {NoStop}%
\bibitem [{\citenamefont {{Ferreira}}(2020)}]{ElisaFerreira}%
  \BibitemOpen
  \bibfield  {author} {\bibinfo {author} {\bibfnamefont {Elisa G.~M.}\
  \bibnamefont {{Ferreira}}},\ }\bibfield  {title} {\enquote {\bibinfo {title}
  {{Ultra-Light Dark Matter}},}\ }\href@noop {} {\bibfield  {journal} {\bibinfo
   {journal} {arXiv e-prints}\ ,\ \bibinfo {eid} {arXiv:2005.03254}} (\bibinfo
  {year} {2020})},\ \Eprint {http://arxiv.org/abs/2005.03254} {arXiv:2005.03254
  [astro-ph.CO]} \BibitemShut {NoStop}%
\bibitem [{\citenamefont {Niemeyer}(2020)}]{Niemeyer_2020}%
  \BibitemOpen
  \bibfield  {author} {\bibinfo {author} {\bibfnamefont {Jens~C.}\ \bibnamefont
  {Niemeyer}},\ }\bibfield  {title} {\enquote {\bibinfo {title} {Small-scale
  structure of fuzzy and axion-like dark matter},}\ }\href {\doibase
  10.1016/j.ppnp.2020.103787} {\bibfield  {journal} {\bibinfo  {journal}
  {Progress in Particle and Nuclear Physics}\ }\textbf {\bibinfo {volume}
  {113}},\ \bibinfo {pages} {103787} (\bibinfo {year} {2020})}\BibitemShut
  {NoStop}%
\bibitem [{\citenamefont {Hui}(2021)}]{Hui:2021tkt}%
  \BibitemOpen
  \bibfield  {author} {\bibinfo {author} {\bibfnamefont {Lam}\ \bibnamefont
  {Hui}},\ }\bibfield  {title} {\enquote {\bibinfo {title} {Wave dark
  matter},}\ }\href {\doibase 10.1146/annurev-astro-120920-010024} {\bibfield
  {journal} {\bibinfo  {journal} {Annual Review of Astronomy and Astrophysics}\
  }\textbf {\bibinfo {volume} {59}},\ \bibinfo {pages} {247--289} (\bibinfo
  {year} {2021})}\BibitemShut {NoStop}%
\bibitem [{\citenamefont {Chavanis}(2025)}]{10.3389/fspas.2025.1538434}%
  \BibitemOpen
  \bibfield  {author} {\bibinfo {author} {\bibfnamefont {Pierre-Henri}\
  \bibnamefont {Chavanis}},\ }\bibfield  {title} {\enquote {\bibinfo {title} {A
  review of basic results on the bose–einstein condensate dark matter
  model},}\ }\href {\doibase 10.3389/fspas.2025.1538434} {\bibfield  {journal}
  {\bibinfo  {journal} {Frontiers in Astronomy and Space Sciences}\ }\textbf
  {\bibinfo {volume} {Volume 12 - 2025}} (\bibinfo {year} {2025}),\
  10.3389/fspas.2025.1538434}\BibitemShut {NoStop}%
\bibitem [{\citenamefont {{Ruffini}}\ and\ \citenamefont
  {{Bonazzola}}(1969)}]{Ruffini:1969}%
  \BibitemOpen
  \bibfield  {author} {\bibinfo {author} {\bibfnamefont {R.}~\bibnamefont
  {{Ruffini}}}\ and\ \bibinfo {author} {\bibfnamefont {S.}~\bibnamefont
  {{Bonazzola}}},\ }\bibfield  {title} {\enquote {\bibinfo {title} {Systems of
  self-gravitating particles in general relativity and the concept of an
  equation of state},}\ }\href {\doibase 10.1103/PhysRev.187.1767} {\bibfield
  {journal} {\bibinfo  {journal} {Phys. Rev.}\ }\textbf {\bibinfo {volume}
  {187}},\ \bibinfo {pages} {1767--1783} (\bibinfo {year} {1969})}\BibitemShut
  {NoStop}%
\bibitem [{\citenamefont {Guzm\'an}\ and\ \citenamefont {Ure\~na
  L\'opez}(2004)}]{GuzmanUrena2004}%
  \BibitemOpen
  \bibfield  {author} {\bibinfo {author} {\bibfnamefont {F.~S.}\ \bibnamefont
  {Guzm\'an}}\ and\ \bibinfo {author} {\bibfnamefont {L.~Arturo}\ \bibnamefont
  {Ure\~na L\'opez}},\ }\bibfield  {title} {\enquote {\bibinfo {title}
  {Evolution of the schr\"odinger-newton system for a self-gravitating scalar
  field},}\ }\href {\doibase 10.1103/PhysRevD.69.124033} {\bibfield  {journal}
  {\bibinfo  {journal} {Phys. Rev. D}\ }\textbf {\bibinfo {volume} {69}},\
  \bibinfo {pages} {124033} (\bibinfo {year} {2004})}\BibitemShut {NoStop}%
\bibitem [{\citenamefont {Guzm\'an}\ and\ \citenamefont {Ure\~na
  L\'opez}(2006)}]{GuzmanUrena2006}%
  \BibitemOpen
  \bibfield  {author} {\bibinfo {author} {\bibfnamefont {F.~S.}\ \bibnamefont
  {Guzm\'an}}\ and\ \bibinfo {author} {\bibfnamefont {L.~Arturo}\ \bibnamefont
  {Ure\~na L\'opez}},\ }\bibfield  {title} {\enquote {\bibinfo {title}
  {Gravitational cooling of self-gravitating bose condensates},}\ }\href
  {\doibase 10.1086/504508} {\bibfield  {journal} {\bibinfo  {journal} {The
  Astrophysical Journal}\ }\textbf {\bibinfo {volume} {645}},\ \bibinfo {pages}
  {814D819} (\bibinfo {year} {2006})}\BibitemShut {NoStop}%
\bibitem [{\citenamefont {Bernal}\ and\ \citenamefont
  {Guzm\'an}(2006)}]{BernalGuzman2006b}%
  \BibitemOpen
  \bibfield  {author} {\bibinfo {author} {\bibfnamefont {Argelia}\ \bibnamefont
  {Bernal}}\ and\ \bibinfo {author} {\bibfnamefont {F.~S.}\ \bibnamefont
  {Guzm\'an}},\ }\bibfield  {title} {\enquote {\bibinfo {title} {Scalar field
  dark matter: Nonspherical collapse and late-time behavior},}\ }\href
  {\doibase 10.1103/physrevd.74.063504} {\bibfield  {journal} {\bibinfo
  {journal} {Physical Review D}\ }\textbf {\bibinfo {volume} {74}} (\bibinfo
  {year} {2006}),\ 10.1103/physrevd.74.063504}\BibitemShut {NoStop}%
\bibitem [{\citenamefont {Schive}\ \emph {et~al.}(2014a)\citenamefont {Schive},
  \citenamefont {Chiueh},\ and\ \citenamefont {Broadhurst}}]{Schive:2014dra}%
  \BibitemOpen
  \bibfield  {author} {\bibinfo {author} {\bibfnamefont {Hsi-Yu}\ \bibnamefont
  {Schive}}, \bibinfo {author} {\bibfnamefont {Tzihong}\ \bibnamefont
  {Chiueh}}, \ and\ \bibinfo {author} {\bibfnamefont {Tom}\ \bibnamefont
  {Broadhurst}},\ }\bibfield  {title} {\enquote {\bibinfo {title} {{Cosmic
  Structure as the Quantum Interference of a Coherent Dark Wave}},}\ }\href
  {\doibase 10.1038/nphys2996} {\bibfield  {journal} {\bibinfo  {journal}
  {Nature Phys.}\ }\textbf {\bibinfo {volume} {10}},\ \bibinfo {pages}
  {496--499} (\bibinfo {year} {2014a})},\ \Eprint
  {http://arxiv.org/abs/1406.6586} {arXiv:1406.6586 [astro-ph.GA]} \BibitemShut
  {NoStop}%
%%CITATION = ARXIV:1406.6586;%%
\bibitem [{\citenamefont {Mocz}\ \emph {et~al.}(2017)\citenamefont {Mocz},
  \citenamefont {Vogelsberger}, \citenamefont {Robles}, \citenamefont {Zavala},
  \citenamefont {Boylan-Kolchin}, \citenamefont {Fialkov},\ and\ \citenamefont
  {Hernquist}}]{Mocz:2017wlg}%
  \BibitemOpen
  \bibfield  {author} {\bibinfo {author} {\bibfnamefont {Philip}\ \bibnamefont
  {Mocz}}, \bibinfo {author} {\bibfnamefont {Mark}\ \bibnamefont
  {Vogelsberger}}, \bibinfo {author} {\bibfnamefont {Victor~H.}\ \bibnamefont
  {Robles}}, \bibinfo {author} {\bibfnamefont {Jes\'us}\ \bibnamefont
  {Zavala}}, \bibinfo {author} {\bibfnamefont {Michael}\ \bibnamefont
  {Boylan-Kolchin}}, \bibinfo {author} {\bibfnamefont {Anastasia}\ \bibnamefont
  {Fialkov}}, \ and\ \bibinfo {author} {\bibfnamefont {Lars}\ \bibnamefont
  {Hernquist}},\ }\bibfield  {title} {\enquote {\bibinfo {title} {Galaxy
  formation with becdm i. turbulence and relaxation of idealized haloes},}\
  }\href {\doibase 10.1093/mnras/stx1887} {\bibfield  {journal} {\bibinfo
  {journal} {Mon. Not. Roy. Astron. Soc.}\ }\textbf {\bibinfo {volume} {471}},\
  \bibinfo {pages} {4559--4570} (\bibinfo {year} {2017})},\ \Eprint
  {http://arxiv.org/abs/1705.05845} {arXiv:1705.05845 [astro-ph.CO]}
  \BibitemShut {NoStop}%
%%CITATION = ARXIV:1705.05845;%%
\bibitem [{\citenamefont {Veltmaat}\ \emph {et~al.}(2018)\citenamefont
  {Veltmaat}, \citenamefont {Niemeyer},\ and\ \citenamefont
  {Schwabe}}]{Veltmaat_2018}%
  \BibitemOpen
  \bibfield  {author} {\bibinfo {author} {\bibfnamefont {Jan}\ \bibnamefont
  {Veltmaat}}, \bibinfo {author} {\bibfnamefont {Jens~C.}\ \bibnamefont
  {Niemeyer}}, \ and\ \bibinfo {author} {\bibfnamefont {Bodo}\ \bibnamefont
  {Schwabe}},\ }\bibfield  {title} {\enquote {\bibinfo {title} {Formation and
  structure of ultralight bosonic dark matter halos},}\ }\href {\doibase
  10.1103/physrevd.98.043509} {\bibfield  {journal} {\bibinfo  {journal}
  {Physical Review D}\ }\textbf {\bibinfo {volume} {98}} (\bibinfo {year}
  {2018}),\ 10.1103/physrevd.98.043509}\BibitemShut {NoStop}%
\bibitem [{\citenamefont {Schwabe}\ and\ \citenamefont
  {Niemeyer}(2022)}]{Gotinga2022}%
  \BibitemOpen
  \bibfield  {author} {\bibinfo {author} {\bibfnamefont {Bodo}\ \bibnamefont
  {Schwabe}}\ and\ \bibinfo {author} {\bibfnamefont {Jens~C.}\ \bibnamefont
  {Niemeyer}},\ }\bibfield  {title} {\enquote {\bibinfo {title} {Deep zoom-in
  simulation of a fuzzy dark matter galactic halo},}\ }\href {\doibase
  10.1103/PhysRevLett.128.181301} {\bibfield  {journal} {\bibinfo  {journal}
  {Phys. Rev. Lett.}\ }\textbf {\bibinfo {volume} {128}},\ \bibinfo {pages}
  {181301} (\bibinfo {year} {2022})}\BibitemShut {NoStop}%
\bibitem [{\citenamefont {Levkov}\ \emph {et~al.}(2018)\citenamefont {Levkov},
  \citenamefont {Panin},\ and\ \citenamefont {Tkachev}}]{Rusos2018}%
  \BibitemOpen
  \bibfield  {author} {\bibinfo {author} {\bibfnamefont {D.~G.}\ \bibnamefont
  {Levkov}}, \bibinfo {author} {\bibfnamefont {A.~G.}\ \bibnamefont {Panin}}, \
  and\ \bibinfo {author} {\bibfnamefont {I.~I.}\ \bibnamefont {Tkachev}},\
  }\bibfield  {title} {\enquote {\bibinfo {title} {Gravitational bose-einstein
  condensation in the kinetic regime},}\ }\href {\doibase
  10.1103/PhysRevLett.121.151301} {\bibfield  {journal} {\bibinfo  {journal}
  {Phys. Rev. Lett.}\ }\textbf {\bibinfo {volume} {121}},\ \bibinfo {pages}
  {151301} (\bibinfo {year} {2018})}\BibitemShut {NoStop}%
\bibitem [{\citenamefont {Eggemeier}\ and\ \citenamefont
  {Niemeyer}(2019)}]{Eggemeier2019}%
  \BibitemOpen
  \bibfield  {author} {\bibinfo {author} {\bibfnamefont {Benedikt}\
  \bibnamefont {Eggemeier}}\ and\ \bibinfo {author} {\bibfnamefont {Jens~C.}\
  \bibnamefont {Niemeyer}},\ }\bibfield  {title} {\enquote {\bibinfo {title}
  {Formation and mass growth of axion stars in axion miniclusters},}\ }\href
  {\doibase 10.1103/PhysRevD.100.063528} {\bibfield  {journal} {\bibinfo
  {journal} {Phys. Rev. D}\ }\textbf {\bibinfo {volume} {100}},\ \bibinfo
  {pages} {063528} (\bibinfo {year} {2019})}\BibitemShut {NoStop}%
\bibitem [{\citenamefont {Chen}\ \emph {et~al.}(2021)\citenamefont {Chen},
  \citenamefont {Du}, \citenamefont {Lentz}, \citenamefont {Marsh},\ and\
  \citenamefont {Niemeyer}}]{Chen2021}%
  \BibitemOpen
  \bibfield  {author} {\bibinfo {author} {\bibfnamefont {Jiajun}\ \bibnamefont
  {Chen}}, \bibinfo {author} {\bibfnamefont {Xiaolong}\ \bibnamefont {Du}},
  \bibinfo {author} {\bibfnamefont {Erik~W.}\ \bibnamefont {Lentz}}, \bibinfo
  {author} {\bibfnamefont {David J.~E.}\ \bibnamefont {Marsh}}, \ and\ \bibinfo
  {author} {\bibfnamefont {Jens~C.}\ \bibnamefont {Niemeyer}},\ }\bibfield
  {title} {\enquote {\bibinfo {title} {New insights into the formation and
  growth of boson stars in dark matter halos},}\ }\href {\doibase
  10.1103/PhysRevD.104.083022} {\bibfield  {journal} {\bibinfo  {journal}
  {Phys. Rev. D}\ }\textbf {\bibinfo {volume} {104}},\ \bibinfo {pages}
  {083022} (\bibinfo {year} {2021})}\BibitemShut {NoStop}%
\bibitem [{\citenamefont {Du}\ \emph {et~al.}(2017)\citenamefont {Du},
  \citenamefont {Behrens}, \citenamefont {Niemeyer},\ and\ \citenamefont
  {Schwabe}}]{DuNiemeyer2017}%
  \BibitemOpen
  \bibfield  {author} {\bibinfo {author} {\bibfnamefont {Xiaolong}\
  \bibnamefont {Du}}, \bibinfo {author} {\bibfnamefont {Christoph}\
  \bibnamefont {Behrens}}, \bibinfo {author} {\bibfnamefont {Jens~C.}\
  \bibnamefont {Niemeyer}}, \ and\ \bibinfo {author} {\bibfnamefont {Bodo}\
  \bibnamefont {Schwabe}},\ }\bibfield  {title} {\enquote {\bibinfo {title}
  {Core-halo mass relation of ultralight axion dark matter from merger
  history},}\ }\href {\doibase 10.1103/PhysRevD.95.043519} {\bibfield
  {journal} {\bibinfo  {journal} {Phys. Rev. D}\ }\textbf {\bibinfo {volume}
  {95}},\ \bibinfo {pages} {043519} (\bibinfo {year} {2017})}\BibitemShut
  {NoStop}%
\bibitem [{\citenamefont {Zagorac}\ \emph {et~al.}(2023)\citenamefont
  {Zagorac}, \citenamefont {Kendall}, \citenamefont {Padmanabhan},\ and\
  \citenamefont {Easther}}]{Luna2023}%
  \BibitemOpen
  \bibfield  {author} {\bibinfo {author} {\bibfnamefont {J.~Luna}\ \bibnamefont
  {Zagorac}}, \bibinfo {author} {\bibfnamefont {Emily}\ \bibnamefont
  {Kendall}}, \bibinfo {author} {\bibfnamefont {Nikhil}\ \bibnamefont
  {Padmanabhan}}, \ and\ \bibinfo {author} {\bibfnamefont {Richard}\
  \bibnamefont {Easther}},\ }\bibfield  {title} {\enquote {\bibinfo {title}
  {Soliton formation and the core-halo mass relation: An eigenstate
  perspective},}\ }\href {\doibase 10.1103/PhysRevD.107.083513} {\bibfield
  {journal} {\bibinfo  {journal} {Phys. Rev. D}\ }\textbf {\bibinfo {volume}
  {107}},\ \bibinfo {pages} {083513} (\bibinfo {year} {2023})}\BibitemShut
  {NoStop}%
\bibitem [{\citenamefont {\'Alvarez-Rios}\ \emph {et~al.}(2023)\citenamefont
  {\'Alvarez-Rios}, \citenamefont {Guzm\'an},\ and\ \citenamefont
  {Shapiro}}]{periodicas}%
  \BibitemOpen
  \bibfield  {author} {\bibinfo {author} {\bibfnamefont {Iv\'an}\ \bibnamefont
  {\'Alvarez-Rios}}, \bibinfo {author} {\bibfnamefont {Francisco~S.}\
  \bibnamefont {Guzm\'an}}, \ and\ \bibinfo {author} {\bibfnamefont {Paul~R.}\
  \bibnamefont {Shapiro}},\ }\bibfield  {title} {\enquote {\bibinfo {title}
  {Effect of boundary conditions on structure formation in fuzzy dark
  matter},}\ }\href {\doibase 10.1103/PhysRevD.107.123524} {\bibfield
  {journal} {\bibinfo  {journal} {Phys. Rev. D}\ }\textbf {\bibinfo {volume}
  {107}},\ \bibinfo {pages} {123524} (\bibinfo {year} {2023})}\BibitemShut
  {NoStop}%
\bibitem [{\citenamefont {Chavanis}(2019)}]{Chavanis_2019}%
  \BibitemOpen
  \bibfield  {author} {\bibinfo {author} {\bibfnamefont {Pierre-Henri}\
  \bibnamefont {Chavanis}},\ }\bibfield  {title} {\enquote {\bibinfo {title}
  {Mass-radius relation of self-gravitating bose-einstein condensates with a
  central black hole},}\ }\href {\doibase 10.1140/epjp/i2019-12734-7}
  {\bibfield  {journal} {\bibinfo  {journal} {Eur. Phys. J. Plus}\ }\textbf
  {\bibinfo {volume} {134}},\ \bibinfo {pages} {352} (\bibinfo {year}
  {2019})}\BibitemShut {NoStop}%
\bibitem [{\citenamefont {Chavanis}(2020)}]{Chavanis_2020}%
  \BibitemOpen
  \bibfield  {author} {\bibinfo {author} {\bibfnamefont {Pierre-Henri}\
  \bibnamefont {Chavanis}},\ }\bibfield  {title} {\enquote {\bibinfo {title}
  {Core mass-halo mass relation of bosonic and fermionic dark matter halos
  harboring a supermassive black hole},}\ }\href {\doibase
  10.1103/physrevd.101.063532} {\bibfield  {journal} {\bibinfo  {journal}
  {Physical Review D}\ }\textbf {\bibinfo {volume} {101}} (\bibinfo {year}
  {2020}),\ 10.1103/physrevd.101.063532}\BibitemShut {NoStop}%
\bibitem [{\citenamefont {\'Alvarez-Rios}\ and\ \citenamefont
  {Guzm\'an}(2023)}]{Alvarez_Rios_2023}%
  \BibitemOpen
  \bibfield  {author} {\bibinfo {author} {\bibfnamefont {Iv\'an}\ \bibnamefont
  {\'Alvarez-Rios}}\ and\ \bibinfo {author} {\bibfnamefont {Francisco~S.}\
  \bibnamefont {Guzm\'an}},\ }\bibfield  {title} {\enquote {\bibinfo {title}
  {Stationary solutions of the schrödinger-poisson-euler system and their
  stability},}\ }\href {\doibase 10.1016/j.physletb.2023.137984} {\bibfield
  {journal} {\bibinfo  {journal} {Physics Letters B}\ }\textbf {\bibinfo
  {volume} {843}},\ \bibinfo {pages} {137984} (\bibinfo {year}
  {2023})}\BibitemShut {NoStop}%
\bibitem [{\citenamefont {Tena-Contreras}\ \emph {et~al.}(2024)\citenamefont
  {Tena-Contreras}, \citenamefont {Alvarez-R\'ios},\ and\ \citenamefont
  {Guzm\'an}}]{CarlosIvanFranciscoUniverse}%
  \BibitemOpen
  \bibfield  {author} {\bibinfo {author} {\bibfnamefont {Carlos}\ \bibnamefont
  {Tena-Contreras}}, \bibinfo {author} {\bibfnamefont {Iv\'an}\ \bibnamefont
  {Alvarez-R\'ios}}, \ and\ \bibinfo {author} {\bibfnamefont {Francisco~S.}\
  \bibnamefont {Guzm\'an}},\ }\bibfield  {title} {\enquote {\bibinfo {title}
  {Construction of ground-state solutions of the gross–pitaevskii–poisson
  system using genetic algorithms},}\ }\href {\doibase
  10.3390/universe10080309} {\bibfield  {journal} {\bibinfo  {journal}
  {Universe}\ }\textbf {\bibinfo {volume} {10}} (\bibinfo {year} {2024}),\
  10.3390/universe10080309}\BibitemShut {NoStop}%
\bibitem [{\citenamefont {Guzm\'an}\ and\ \citenamefont {Ure\~na
  L\'opez}(2020)}]{GuzmanUrena2020}%
  \BibitemOpen
  \bibfield  {author} {\bibinfo {author} {\bibfnamefont {F.~S.}\ \bibnamefont
  {Guzm\'an}}\ and\ \bibinfo {author} {\bibfnamefont {L.~Arturo}\ \bibnamefont
  {Ure\~na L\'opez}},\ }\bibfield  {title} {\enquote {\bibinfo {title}
  {Gravitational atoms: General framework for the construction of multistate
  axially symmetric solutions of the schr\"odinger-poisson system},}\ }\href
  {\doibase 10.1103/PhysRevD.101.081302} {\bibfield  {journal} {\bibinfo
  {journal} {Phys. Rev. D}\ }\textbf {\bibinfo {volume} {101}},\ \bibinfo
  {pages} {081302} (\bibinfo {year} {2020})}\BibitemShut {NoStop}%
\bibitem [{\citenamefont {Sol\'{\i}s-L\'opez}\ \emph
  {et~al.}(2021)\citenamefont {Sol\'{\i}s-L\'opez}, \citenamefont {Guzm\'an},
  \citenamefont {Matos}, \citenamefont {Robles},\ and\ \citenamefont {Ure\~na
  L\'opez}}]{jordi}%
  \BibitemOpen
  \bibfield  {author} {\bibinfo {author} {\bibfnamefont {Jordi}\ \bibnamefont
  {Sol\'{\i}s-L\'opez}}, \bibinfo {author} {\bibfnamefont {Francisco~S.}\
  \bibnamefont {Guzm\'an}}, \bibinfo {author} {\bibfnamefont {Tonatiuh}\
  \bibnamefont {Matos}}, \bibinfo {author} {\bibfnamefont {Victor~H.}\
  \bibnamefont {Robles}}, \ and\ \bibinfo {author} {\bibfnamefont {L.~Arturo}\
  \bibnamefont {Ure\~na L\'opez}},\ }\bibfield  {title} {\enquote {\bibinfo
  {title} {Scalar field dark matter as an alternative explanation for the
  anisotropic distribution of satellite galaxies},}\ }\href {\doibase
  10.1103/PhysRevD.103.083535} {\bibfield  {journal} {\bibinfo  {journal}
  {Phys. Rev. D}\ }\textbf {\bibinfo {volume} {103}},\ \bibinfo {pages}
  {083535} (\bibinfo {year} {2021})}\BibitemShut {NoStop}%
\bibitem [{\citenamefont {Matthews}\ \emph {et~al.}(1999)\citenamefont
  {Matthews}, \citenamefont {Anderson}, \citenamefont {Haljan}, \citenamefont
  {Hall}, \citenamefont {Wieman},\ and\ \citenamefont
  {Cornell}}]{Matthews1999}%
  \BibitemOpen
  \bibfield  {author} {\bibinfo {author} {\bibfnamefont {M.~R.}\ \bibnamefont
  {Matthews}}, \bibinfo {author} {\bibfnamefont {B.~P.}\ \bibnamefont
  {Anderson}}, \bibinfo {author} {\bibfnamefont {P.~C.}\ \bibnamefont
  {Haljan}}, \bibinfo {author} {\bibfnamefont {D.~S.}\ \bibnamefont {Hall}},
  \bibinfo {author} {\bibfnamefont {C.~E.}\ \bibnamefont {Wieman}}, \ and\
  \bibinfo {author} {\bibfnamefont {E.~A.}\ \bibnamefont {Cornell}},\
  }\bibfield  {title} {\enquote {\bibinfo {title} {Vortices in a bose-einstein
  condensate},}\ }\href {\doibase 10.1103/PhysRevLett.83.2498} {\bibfield
  {journal} {\bibinfo  {journal} {Phys. Rev. Lett.}\ }\textbf {\bibinfo
  {volume} {83}},\ \bibinfo {pages} {2498--2501} (\bibinfo {year}
  {1999})}\BibitemShut {NoStop}%
\bibitem [{\citenamefont {Leanhardt}\ \emph {et~al.}(2002)\citenamefont
  {Leanhardt}, \citenamefont {G\"orlitz}, \citenamefont {Chikkatur},
  \citenamefont {Kielpinski}, \citenamefont {Shin}, \citenamefont {Pritchard},\
  and\ \citenamefont {Ketterle}}]{Leanhardt2002}%
  \BibitemOpen
  \bibfield  {author} {\bibinfo {author} {\bibfnamefont {A.~E.}\ \bibnamefont
  {Leanhardt}}, \bibinfo {author} {\bibfnamefont {A.}~\bibnamefont
  {G\"orlitz}}, \bibinfo {author} {\bibfnamefont {A.~P.}\ \bibnamefont
  {Chikkatur}}, \bibinfo {author} {\bibfnamefont {D.}~\bibnamefont
  {Kielpinski}}, \bibinfo {author} {\bibfnamefont {Y.}~\bibnamefont {Shin}},
  \bibinfo {author} {\bibfnamefont {D.~E.}\ \bibnamefont {Pritchard}}, \ and\
  \bibinfo {author} {\bibfnamefont {W.}~\bibnamefont {Ketterle}},\ }\bibfield
  {title} {\enquote {\bibinfo {title} {Imprinting vortices in a bose-einstein
  condensate using topological phases},}\ }\href {\doibase
  10.1103/PhysRevLett.89.190403} {\bibfield  {journal} {\bibinfo  {journal}
  {Phys. Rev. Lett.}\ }\textbf {\bibinfo {volume} {89}},\ \bibinfo {pages}
  {190403} (\bibinfo {year} {2002})}\BibitemShut {NoStop}%
\bibitem [{\citenamefont {Fetter}(2009)}]{Fetter2009}%
  \BibitemOpen
  \bibfield  {author} {\bibinfo {author} {\bibfnamefont {Alexander~L.}\
  \bibnamefont {Fetter}},\ }\bibfield  {title} {\enquote {\bibinfo {title}
  {Rotating trapped bose-einstein condensates},}\ }\href {\doibase
  10.1103/RevModPhys.81.647} {\bibfield  {journal} {\bibinfo  {journal} {Rev.
  Mod. Phys.}\ }\textbf {\bibinfo {volume} {81}},\ \bibinfo {pages} {647--691}
  (\bibinfo {year} {2009})}\BibitemShut {NoStop}%
\bibitem [{\citenamefont {Pethick}\ and\ \citenamefont
  {Smith}(2008)}]{PethickSmith}%
  \BibitemOpen
  \bibfield  {author} {\bibinfo {author} {\bibfnamefont {C.~J.}\ \bibnamefont
  {Pethick}}\ and\ \bibinfo {author} {\bibfnamefont {H.}~\bibnamefont
  {Smith}},\ }\href@noop {} {\emph {\bibinfo {title} {Bose–Einstein
  Condensation in Dilute Gases}}},\ \bibinfo {edition} {2nd}\ ed.\ (\bibinfo
  {publisher} {Cambridge University Press},\ \bibinfo {year}
  {2008})\BibitemShut {NoStop}%
\bibitem [{\citenamefont {Rindler-Daller}\ and\ \citenamefont
  {Shapiro}(2012)}]{Rindler_Daller_2012}%
  \BibitemOpen
  \bibfield  {author} {\bibinfo {author} {\bibfnamefont {Tanja}\ \bibnamefont
  {Rindler-Daller}}\ and\ \bibinfo {author} {\bibfnamefont {Paul~R.}\
  \bibnamefont {Shapiro}},\ }\bibfield  {title} {\enquote {\bibinfo {title}
  {Angular momentum and vortex formation in bose-einstein-condensed cold dark
  matter haloes: Angular momentum in bec-cdm haloes},}\ }\href {\doibase
  10.1111/j.1365-2966.2012.20588.x} {\bibfield  {journal} {\bibinfo  {journal}
  {Monthly Notices of the Royal Astronomical Society}\ }\textbf {\bibinfo
  {volume} {422}},\ \bibinfo {pages} {135–161} (\bibinfo {year}
  {2012})}\BibitemShut {NoStop}%
\bibitem [{\citenamefont {Kain}\ and\ \citenamefont {Ling}(2010)}]{Kain_2010}%
  \BibitemOpen
  \bibfield  {author} {\bibinfo {author} {\bibfnamefont {Ben}\ \bibnamefont
  {Kain}}\ and\ \bibinfo {author} {\bibfnamefont {Hong~Y.}\ \bibnamefont
  {Ling}},\ }\bibfield  {title} {\enquote {\bibinfo {title} {Vortices in
  bose-einstein condensate dark matter},}\ }\href {\doibase
  10.1103/physrevd.82.064042} {\bibfield  {journal} {\bibinfo  {journal}
  {Physical Review D}\ }\textbf {\bibinfo {volume} {82}} (\bibinfo {year}
  {2010}),\ 10.1103/physrevd.82.064042}\BibitemShut {NoStop}%
\bibitem [{\citenamefont {Korshynska}\ \emph {et~al.}(2023)\citenamefont
  {Korshynska}, \citenamefont {Bidasyuk}, \citenamefont {Gorbar}, \citenamefont
  {Jia},\ and\ \citenamefont {Yakimenko}}]{Korshynska_2023}%
  \BibitemOpen
  \bibfield  {author} {\bibinfo {author} {\bibfnamefont {K.}~\bibnamefont
  {Korshynska}}, \bibinfo {author} {\bibfnamefont {Y.~M.}\ \bibnamefont
  {Bidasyuk}}, \bibinfo {author} {\bibfnamefont {E.~V.}\ \bibnamefont
  {Gorbar}}, \bibinfo {author} {\bibfnamefont {Junji}\ \bibnamefont {Jia}}, \
  and\ \bibinfo {author} {\bibfnamefont {A.~I.}\ \bibnamefont {Yakimenko}},\
  }\bibfield  {title} {\enquote {\bibinfo {title} {Dynamical galactic effects
  induced by solitonic vortex structure in bosonic dark matter},}\ }\href
  {\doibase 10.1140/epjc/s10052-023-11548-1} {\bibfield  {journal} {\bibinfo
  {journal} {The European Physical Journal C}\ }\textbf {\bibinfo {volume}
  {83}} (\bibinfo {year} {2023}),\ 10.1140/epjc/s10052-023-11548-1}\BibitemShut
  {NoStop}%
\bibitem [{\citenamefont {Nikolaieva}\ \emph {et~al.}(2023)\citenamefont
  {Nikolaieva}, \citenamefont {Bidasyuk}, \citenamefont {Korshynska},
  \citenamefont {Gorbar}, \citenamefont {Jia},\ and\ \citenamefont
  {Yakimenko}}]{Nikolaieva_2023}%
  \BibitemOpen
  \bibfield  {author} {\bibinfo {author} {\bibfnamefont {Y.~O.}\ \bibnamefont
  {Nikolaieva}}, \bibinfo {author} {\bibfnamefont {Y.~M.}\ \bibnamefont
  {Bidasyuk}}, \bibinfo {author} {\bibfnamefont {K.}~\bibnamefont
  {Korshynska}}, \bibinfo {author} {\bibfnamefont {E.~V.}\ \bibnamefont
  {Gorbar}}, \bibinfo {author} {\bibfnamefont {Junji}\ \bibnamefont {Jia}}, \
  and\ \bibinfo {author} {\bibfnamefont {A.~I.}\ \bibnamefont {Yakimenko}},\
  }\bibfield  {title} {\enquote {\bibinfo {title} {Stable vortex structures in
  colliding self-gravitating bose-einstein condensates},}\ }\href {\doibase
  10.1103/physrevd.108.023503} {\bibfield  {journal} {\bibinfo  {journal}
  {Physical Review D}\ }\textbf {\bibinfo {volume} {108}} (\bibinfo {year}
  {2023}),\ 10.1103/physrevd.108.023503}\BibitemShut {NoStop}%
\bibitem [{\citenamefont {Alexander}\ \emph {et~al.}(2022)\citenamefont
  {Alexander}, \citenamefont {Capanelli}, \citenamefont {G.~M.~Ferreira},\ and\
  \citenamefont {McDonough}}]{Alexander:2021zhx}%
  \BibitemOpen
  \bibfield  {author} {\bibinfo {author} {\bibfnamefont {Stephon}\ \bibnamefont
  {Alexander}}, \bibinfo {author} {\bibfnamefont {Christian}\ \bibnamefont
  {Capanelli}}, \bibinfo {author} {\bibfnamefont {Elisa}\ \bibnamefont
  {G.~M.~Ferreira}}, \ and\ \bibinfo {author} {\bibfnamefont {Evan}\
  \bibnamefont {McDonough}},\ }\bibfield  {title} {\enquote {\bibinfo {title}
  {{Cosmic filament spin from dark matter vortices}},}\ }\href {\doibase
  10.1016/j.physletb.2022.137298} {\bibfield  {journal} {\bibinfo  {journal}
  {Phys. Lett. B}\ }\textbf {\bibinfo {volume} {833}},\ \bibinfo {pages}
  {137298} (\bibinfo {year} {2022})},\ \Eprint
  {http://arxiv.org/abs/2111.03061} {arXiv:2111.03061 [astro-ph.CO]}
  \BibitemShut {NoStop}%
\bibitem [{\citenamefont {Álvarez Rios}\ \emph {et~al.}(2025)\citenamefont
  {Álvarez Rios}, \citenamefont {Tena-Contreras},\ and\ \citenamefont
  {Guzmán}}]{Alvarez_Rios_2025}%
  \BibitemOpen
  \bibfield  {author} {\bibinfo {author} {\bibfnamefont {Iván}\ \bibnamefont
  {Álvarez Rios}}, \bibinfo {author} {\bibfnamefont {Carlos}\ \bibnamefont
  {Tena-Contreras}}, \ and\ \bibinfo {author} {\bibfnamefont {Francisco~S.}\
  \bibnamefont {Guzmán}},\ }\bibfield  {title} {\enquote {\bibinfo {title}
  {Kinematic imprints of vortex lines of bec dark matter on baryonic matter},}\
  }\href {\doibase 10.1103/x9mt-wprk} {\bibfield  {journal} {\bibinfo
  {journal} {Physical Review D}\ }\textbf {\bibinfo {volume} {111}} (\bibinfo
  {year} {2025}),\ 10.1103/x9mt-wprk}\BibitemShut {NoStop}%
\bibitem [{\citenamefont {Korshynska}\ \emph {et~al.}(2025)\citenamefont
  {Korshynska}, \citenamefont {Prykhodko}, \citenamefont {Gorbar},
  \citenamefont {Jia},\ and\ \citenamefont {Yakimenko}}]{Korshynska_2025}%
  \BibitemOpen
  \bibfield  {author} {\bibinfo {author} {\bibfnamefont {K.}~\bibnamefont
  {Korshynska}}, \bibinfo {author} {\bibfnamefont {O.~O.}\ \bibnamefont
  {Prykhodko}}, \bibinfo {author} {\bibfnamefont {E.~V.}\ \bibnamefont
  {Gorbar}}, \bibinfo {author} {\bibfnamefont {Junji}\ \bibnamefont {Jia}}, \
  and\ \bibinfo {author} {\bibfnamefont {A.~I.}\ \bibnamefont {Yakimenko}},\
  }\bibfield  {title} {\enquote {\bibinfo {title} {Vortex lines in ultralight
  bosonic dark matter around rotating supermassive black holes},}\ }\href
  {\doibase 10.1103/physrevd.111.023006} {\bibfield  {journal} {\bibinfo
  {journal} {Physical Review D}\ }\textbf {\bibinfo {volume} {111}} (\bibinfo
  {year} {2025}),\ 10.1103/physrevd.111.023006}\BibitemShut {NoStop}%
\bibitem [{\citenamefont {Glennon}\ \emph
  {et~al.}(2023{\natexlab{a}})\citenamefont {Glennon}, \citenamefont
  {Mirasola}, \citenamefont {Musoke}, \citenamefont {Neyrinck},\ and\
  \citenamefont {Prescod-Weinstein}}]{Glennon2023}%
  \BibitemOpen
  \bibfield  {author} {\bibinfo {author} {\bibfnamefont {Noah}\ \bibnamefont
  {Glennon}}, \bibinfo {author} {\bibfnamefont {Anthony~E.}\ \bibnamefont
  {Mirasola}}, \bibinfo {author} {\bibfnamefont {Nathan}\ \bibnamefont
  {Musoke}}, \bibinfo {author} {\bibfnamefont {Mark~C.}\ \bibnamefont
  {Neyrinck}}, \ and\ \bibinfo {author} {\bibfnamefont {Chanda}\ \bibnamefont
  {Prescod-Weinstein}},\ }\bibfield  {title} {\enquote {\bibinfo {title}
  {Scalar dark matter vortex stabilization with black holes},}\ }\href
  {\doibase 10.1088/1475-7516/2023/07/004} {\bibfield  {journal} {\bibinfo
  {journal} {Journal of Cosmology and Astroparticle Physics}\ }\textbf
  {\bibinfo {volume} {2023}},\ \bibinfo {pages} {004} (\bibinfo {year}
  {2023}{\natexlab{a}})}\BibitemShut {NoStop}%
\bibitem [{\citenamefont {Ghez}\ \emph {et~al.}(2008)\citenamefont {Ghez},
  \citenamefont {Salim}, \citenamefont {Weinberg} \emph {et~al.}}]{Ghez2008}%
  \BibitemOpen
  \bibfield  {author} {\bibinfo {author} {\bibfnamefont {A.~M.}\ \bibnamefont
  {Ghez}}, \bibinfo {author} {\bibfnamefont {S.}~\bibnamefont {Salim}},
  \bibinfo {author} {\bibfnamefont {N.~N.}\ \bibnamefont {Weinberg}},  \emph
  {et~al.},\ }\bibfield  {title} {\enquote {\bibinfo {title} {Measuring
  distance and properties of the milky way's central supermassive black hole
  with stellar orbits},}\ }\href {\doibase 10.1086/592738} {\bibfield
  {journal} {\bibinfo  {journal} {Astrophys. J.}\ }\textbf {\bibinfo {volume}
  {689}},\ \bibinfo {pages} {1044--1062} (\bibinfo {year} {2008})},\ \Eprint
  {http://arxiv.org/abs/0808.2870} {arXiv:0808.2870 [astro-ph]} \BibitemShut
  {NoStop}%
\bibitem [{\citenamefont {Gillessen}\ \emph {et~al.}(2009)\citenamefont
  {Gillessen}, \citenamefont {Eisenhauer}, \citenamefont {Trippe} \emph
  {et~al.}}]{Gillessen2009}%
  \BibitemOpen
  \bibfield  {author} {\bibinfo {author} {\bibfnamefont {S.}~\bibnamefont
  {Gillessen}}, \bibinfo {author} {\bibfnamefont {F.}~\bibnamefont
  {Eisenhauer}}, \bibinfo {author} {\bibfnamefont {S.}~\bibnamefont {Trippe}},
  \emph {et~al.},\ }\bibfield  {title} {\enquote {\bibinfo {title} {Monitoring
  stellar orbits around the massive black hole in the galactic center},}\
  }\href {\doibase 10.1088/0004-637X/692/2/1075} {\bibfield  {journal}
  {\bibinfo  {journal} {Astrophys. J.}\ }\textbf {\bibinfo {volume} {692}},\
  \bibinfo {pages} {1075--1109} (\bibinfo {year} {2009})},\ \Eprint
  {http://arxiv.org/abs/0810.4674} {arXiv:0810.4674 [astro-ph]} \BibitemShut
  {NoStop}%
\bibitem [{\citenamefont {{GRAVITY Collaboration}}\ \emph
  {et~al.}(2019)\citenamefont {{GRAVITY Collaboration}}, \citenamefont
  {Abuter}, \citenamefont {Amorim} \emph {et~al.}}]{Gravity2019}%
  \BibitemOpen
  \bibfield  {author} {\bibinfo {author} {\bibnamefont {{GRAVITY
  Collaboration}}}, \bibinfo {author} {\bibfnamefont {R.}~\bibnamefont
  {Abuter}}, \bibinfo {author} {\bibfnamefont {A.}~\bibnamefont {Amorim}},
  \emph {et~al.},\ }\bibfield  {title} {\enquote {\bibinfo {title} {A geometric
  distance measurement to the galactic center black hole with 0.3\%
  uncertainty},}\ }\href {\doibase 10.1051/0004-6361/201935656} {\bibfield
  {journal} {\bibinfo  {journal} {Astron. Astrophys.}\ }\textbf {\bibinfo
  {volume} {625}},\ \bibinfo {pages} {L10} (\bibinfo {year} {2019})},\ \Eprint
  {http://arxiv.org/abs/1904.05721} {arXiv:1904.05721 [astro-ph.GA]}
  \BibitemShut {NoStop}%
\bibitem [{\citenamefont {Hertzberg}\ \emph {et~al.}(2020)\citenamefont
  {Hertzberg}, \citenamefont {Schiappacasse},\ and\ \citenamefont
  {Yanagida}}]{Hertzberg2020}%
  \BibitemOpen
  \bibfield  {author} {\bibinfo {author} {\bibfnamefont {Mark~P.}\ \bibnamefont
  {Hertzberg}}, \bibinfo {author} {\bibfnamefont {Enrico~D.}\ \bibnamefont
  {Schiappacasse}}, \ and\ \bibinfo {author} {\bibfnamefont {Tsutomu~T.}\
  \bibnamefont {Yanagida}},\ }\bibfield  {title} {\enquote {\bibinfo {title}
  {Axion star nucleation in dark minihalos around primordial black holes},}\
  }\href {\doibase 10.1103/PhysRevD.102.023013} {\bibfield  {journal} {\bibinfo
   {journal} {Phys. Rev. D}\ }\textbf {\bibinfo {volume} {102}},\ \bibinfo
  {pages} {023013} (\bibinfo {year} {2020})}\BibitemShut {NoStop}%
\bibitem [{\citenamefont {Cardoso}\ \emph {et~al.}(2022)\citenamefont
  {Cardoso}, \citenamefont {Ikeda}, \citenamefont {Vicente},\ and\
  \citenamefont {Zilhão}}]{Cardoso2022}%
  \BibitemOpen
  \bibfield  {author} {\bibinfo {author} {\bibfnamefont {Vitor}\ \bibnamefont
  {Cardoso}}, \bibinfo {author} {\bibfnamefont {Taishi}\ \bibnamefont {Ikeda}},
  \bibinfo {author} {\bibfnamefont {Rodrigo}\ \bibnamefont {Vicente}}, \ and\
  \bibinfo {author} {\bibfnamefont {Miguel}\ \bibnamefont {Zilhão}},\
  }\bibfield  {title} {\enquote {\bibinfo {title} {Parasitic black holes: The
  swallowing of a fuzzy dark matter soliton},}\ }\href {\doibase
  10.1103/physrevd.106.l121302} {\bibfield  {journal} {\bibinfo  {journal}
  {Physical Review D}\ }\textbf {\bibinfo {volume} {106}} (\bibinfo {year}
  {2022}),\ 10.1103/physrevd.106.l121302}\BibitemShut {NoStop}%
\bibitem [{\citenamefont {Boudon}\ \emph {et~al.}(2023)\citenamefont {Boudon},
  \citenamefont {Brax},\ and\ \citenamefont {Valageas}}]{Boudon2023}%
  \BibitemOpen
  \bibfield  {author} {\bibinfo {author} {\bibfnamefont {Alexis}\ \bibnamefont
  {Boudon}}, \bibinfo {author} {\bibfnamefont {Philippe}\ \bibnamefont {Brax}},
  \ and\ \bibinfo {author} {\bibfnamefont {Patrick}\ \bibnamefont {Valageas}},\
  }\bibfield  {title} {\enquote {\bibinfo {title} {Supersonic friction of a
  black hole traversing a self-interacting scalar dark matter cloud},}\ }\href
  {\doibase 10.1103/PhysRevD.108.103517} {\bibfield  {journal} {\bibinfo
  {journal} {Phys. Rev. D}\ }\textbf {\bibinfo {volume} {108}},\ \bibinfo
  {pages} {103517} (\bibinfo {year} {2023})}\BibitemShut {NoStop}%
\bibitem [{\citenamefont {Ravanal}\ \emph {et~al.}(2023)\citenamefont
  {Ravanal}, \citenamefont {Gómez},\ and\ \citenamefont {Cruz}}]{Ravanal2023}%
  \BibitemOpen
  \bibfield  {author} {\bibinfo {author} {\bibfnamefont {Yuri}\ \bibnamefont
  {Ravanal}}, \bibinfo {author} {\bibfnamefont {Gabriel}\ \bibnamefont
  {Gómez}}, \ and\ \bibinfo {author} {\bibfnamefont {Normal}\ \bibnamefont
  {Cruz}},\ }\bibfield  {title} {\enquote {\bibinfo {title} {Accretion of
  self-interacting scalar field dark matter onto a reissner-nordström black
  hole},}\ }\href {\doibase 10.1103/physrevd.108.083004} {\bibfield  {journal}
  {\bibinfo  {journal} {Physical Review D}\ }\textbf {\bibinfo {volume} {108}}
  (\bibinfo {year} {2023}),\ 10.1103/physrevd.108.083004}\BibitemShut {NoStop}%
\bibitem [{\citenamefont {Wang}\ and\ \citenamefont
  {Easther}(2022)}]{Wang2022}%
  \BibitemOpen
  \bibfield  {author} {\bibinfo {author} {\bibfnamefont {Yourong}\ \bibnamefont
  {Wang}}\ and\ \bibinfo {author} {\bibfnamefont {Richard}\ \bibnamefont
  {Easther}},\ }\bibfield  {title} {\enquote {\bibinfo {title} {Dynamical
  friction from ultralight dark matter},}\ }\href {\doibase
  10.1103/PhysRevD.105.063523} {\bibfield  {journal} {\bibinfo  {journal}
  {Phys. Rev. D}\ }\textbf {\bibinfo {volume} {105}},\ \bibinfo {pages}
  {063523} (\bibinfo {year} {2022})}\BibitemShut {NoStop}%
\bibitem [{\citenamefont {Palomares-Chávez}\ \emph {et~al.}(2025)\citenamefont
  {Palomares-Chávez}, \citenamefont {Álvarez Rios},\ and\ \citenamefont
  {Guzmán}}]{palomareschavez2025}%
  \BibitemOpen
  \bibfield  {author} {\bibinfo {author} {\bibfnamefont {Curicaveri}\
  \bibnamefont {Palomares-Chávez}}, \bibinfo {author} {\bibfnamefont {Iván}\
  \bibnamefont {Álvarez Rios}}, \ and\ \bibinfo {author} {\bibfnamefont
  {Francisco~S.}\ \bibnamefont {Guzmán}},\ }\bibfield  {title} {\enquote
  {\bibinfo {title} {Black holes as condensation points of fuzzy dark matter
  cores},}\ }\href {\doibase 10.1103/fwf5-n21g} {\bibfield  {journal} {\bibinfo
   {journal} {Physical Review D}\ }\textbf {\bibinfo {volume} {112}} (\bibinfo
  {year} {2025}),\ 10.1103/fwf5-n21g}\BibitemShut {NoStop}%
\bibitem [{\citenamefont {Bamber}\ \emph {et~al.}(2023)\citenamefont {Bamber},
  \citenamefont {Aurrekoetxea}, \citenamefont {Clough},\ and\ \citenamefont
  {Ferreira}}]{Bamber2023}%
  \BibitemOpen
  \bibfield  {author} {\bibinfo {author} {\bibfnamefont {Jamie}\ \bibnamefont
  {Bamber}}, \bibinfo {author} {\bibfnamefont {Josu~C.}\ \bibnamefont
  {Aurrekoetxea}}, \bibinfo {author} {\bibfnamefont {Katy}\ \bibnamefont
  {Clough}}, \ and\ \bibinfo {author} {\bibfnamefont {Pedro~G.}\ \bibnamefont
  {Ferreira}},\ }\bibfield  {title} {\enquote {\bibinfo {title} {Black hole
  merger simulations in wave dark matter environments},}\ }\href {\doibase
  10.1103/physrevd.107.024035} {\bibfield  {journal} {\bibinfo  {journal}
  {Physical Review D}\ }\textbf {\bibinfo {volume} {107}} (\bibinfo {year}
  {2023}),\ 10.1103/physrevd.107.024035}\BibitemShut {NoStop}%
\bibitem [{\citenamefont {Aurrekoetxea}\ \emph
  {et~al.}(2024{\natexlab{a}})\citenamefont {Aurrekoetxea}, \citenamefont
  {Clough}, \citenamefont {Bamber},\ and\ \citenamefont
  {Ferreira}}]{Aurrekoetxea2024}%
  \BibitemOpen
  \bibfield  {author} {\bibinfo {author} {\bibfnamefont {Josu~C.}\ \bibnamefont
  {Aurrekoetxea}}, \bibinfo {author} {\bibfnamefont {Katy}\ \bibnamefont
  {Clough}}, \bibinfo {author} {\bibfnamefont {Jamie}\ \bibnamefont {Bamber}},
  \ and\ \bibinfo {author} {\bibfnamefont {Pedro~G.}\ \bibnamefont
  {Ferreira}},\ }\bibfield  {title} {\enquote {\bibinfo {title} {Effect of wave
  dark matter on equal mass black hole mergers},}\ }\href {\doibase
  10.1103/physrevlett.132.211401} {\bibfield  {journal} {\bibinfo  {journal}
  {Physical Review Letters}\ }\textbf {\bibinfo {volume} {132}} (\bibinfo
  {year} {2024}{\natexlab{a}}),\ 10.1103/physrevlett.132.211401}\BibitemShut
  {NoStop}%
\bibitem [{\citenamefont {Aurrekoetxea}\ \emph
  {et~al.}(2024{\natexlab{b}})\citenamefont {Aurrekoetxea}, \citenamefont
  {Marsden}, \citenamefont {Clough},\ and\ \citenamefont
  {Ferreira}}]{Aurrekoetxea_2024_selfinteracting}%
  \BibitemOpen
  \bibfield  {author} {\bibinfo {author} {\bibfnamefont {Josu~C.}\ \bibnamefont
  {Aurrekoetxea}}, \bibinfo {author} {\bibfnamefont {James}\ \bibnamefont
  {Marsden}}, \bibinfo {author} {\bibfnamefont {Katy}\ \bibnamefont {Clough}},
  \ and\ \bibinfo {author} {\bibfnamefont {Pedro~G.}\ \bibnamefont
  {Ferreira}},\ }\bibfield  {title} {\enquote {\bibinfo {title}
  {Self-interacting scalar dark matter around binary black holes},}\ }\href
  {\doibase 10.1103/physrevd.110.083011} {\bibfield  {journal} {\bibinfo
  {journal} {Physical Review D}\ }\textbf {\bibinfo {volume} {110}} (\bibinfo
  {year} {2024}{\natexlab{b}}),\ 10.1103/physrevd.110.083011}\BibitemShut
  {NoStop}%
\bibitem [{\citenamefont {Bromley}\ \emph {et~al.}(2024)\citenamefont
  {Bromley}, \citenamefont {Sandick},\ and\ \citenamefont {Shams
  Es~Haghi}}]{Bromley2024}%
  \BibitemOpen
  \bibfield  {author} {\bibinfo {author} {\bibfnamefont {Benjamin~C.}\
  \bibnamefont {Bromley}}, \bibinfo {author} {\bibfnamefont {Pearl}\
  \bibnamefont {Sandick}}, \ and\ \bibinfo {author} {\bibfnamefont {Barmak}\
  \bibnamefont {Shams Es~Haghi}},\ }\bibfield  {title} {\enquote {\bibinfo
  {title} {Supermassive black hole binaries in ultralight dark matter},}\
  }\href {\doibase 10.1103/PhysRevD.110.023517} {\bibfield  {journal} {\bibinfo
   {journal} {Phys. Rev. D}\ }\textbf {\bibinfo {volume} {110}},\ \bibinfo
  {pages} {023517} (\bibinfo {year} {2024})}\BibitemShut {NoStop}%
\bibitem [{\citenamefont {Chavanis}(2011)}]{Chavanis:2011}%
  \BibitemOpen
  \bibfield  {author} {\bibinfo {author} {\bibfnamefont {P.-H.}\ \bibnamefont
  {Chavanis}},\ }\bibfield  {title} {\enquote {\bibinfo {title} {{Mass-radius
  relation of Newtonian self-gravitating Bose-Einstein condensates with
  short-range interactions: I. Analytical results}},}\ }\href {\doibase
  10.1103/PhysRevD.84.043531} {\bibfield  {journal} {\bibinfo  {journal} {Phys.
  Rev. D}\ }\textbf {\bibinfo {volume} {84}},\ \bibinfo {pages} {043531}
  (\bibinfo {year} {2011})},\ \Eprint {http://arxiv.org/abs/1103.2050}
  {arXiv:1103.2050} \BibitemShut {NoStop}%
\bibitem [{\citenamefont {Chavanis}\ and\ \citenamefont
  {Delfini}(2011)}]{PhysRevD.84.043532}%
  \BibitemOpen
  \bibfield  {author} {\bibinfo {author} {\bibfnamefont {Pierre-Henri}\
  \bibnamefont {Chavanis}}\ and\ \bibinfo {author} {\bibfnamefont {Luca}\
  \bibnamefont {Delfini}},\ }\bibfield  {title} {\enquote {\bibinfo {title}
  {Mass-radius relation of newtonian self-gravitating bose-einstein condensates
  with short-range interactions. ii. numerical results},}\ }\href {\doibase
  10.1103/PhysRevD.84.043532} {\bibfield  {journal} {\bibinfo  {journal} {Phys.
  Rev. D}\ }\textbf {\bibinfo {volume} {84}},\ \bibinfo {pages} {043532}
  (\bibinfo {year} {2011})}\BibitemShut {NoStop}%
\bibitem [{\citenamefont {Vakhitov}\ and\ \citenamefont
  {Kolokolov}(1973)}]{VakhitovKolokolov1973}%
  \BibitemOpen
  \bibfield  {author} {\bibinfo {author} {\bibfnamefont {N.~G.}\ \bibnamefont
  {Vakhitov}}\ and\ \bibinfo {author} {\bibfnamefont {A.~A.}\ \bibnamefont
  {Kolokolov}},\ }\bibfield  {title} {\enquote {\bibinfo {title} {Stationary
  solutions of the wave equation in a medium with nonlinearity saturation},}\
  }\href {\doibase 10.1007/BF01031343} {\bibfield  {journal} {\bibinfo
  {journal} {Radiophysics and Quantum Electronics}\ }\textbf {\bibinfo {volume}
  {16}},\ \bibinfo {pages} {783--789} (\bibinfo {year} {1973})}\BibitemShut
  {NoStop}%
\bibitem [{\citenamefont {Nikolaieva}\ \emph {et~al.}(2021)\citenamefont
  {Nikolaieva}, \citenamefont {Olashyn}, \citenamefont {Kuriatnikov},
  \citenamefont {Vilchynskii},\ and\ \citenamefont
  {Yakimenko}}]{Nikolaieva_2021}%
  \BibitemOpen
  \bibfield  {author} {\bibinfo {author} {\bibfnamefont {Y.~O.}\ \bibnamefont
  {Nikolaieva}}, \bibinfo {author} {\bibfnamefont {A.~O.}\ \bibnamefont
  {Olashyn}}, \bibinfo {author} {\bibfnamefont {Y.~I.}\ \bibnamefont
  {Kuriatnikov}}, \bibinfo {author} {\bibfnamefont {S.~I.}\ \bibnamefont
  {Vilchynskii}}, \ and\ \bibinfo {author} {\bibfnamefont {A.~I.}\ \bibnamefont
  {Yakimenko}},\ }\bibfield  {title} {\enquote {\bibinfo {title} {Stable vortex
  in bose-einstein condensate dark matter},}\ }\href {\doibase
  10.1063/10.0005557} {\bibfield  {journal} {\bibinfo  {journal} {Low
  Temperature Physics}\ }\textbf {\bibinfo {volume} {47}},\ \bibinfo {pages}
  {684–692} (\bibinfo {year} {2021})}\BibitemShut {NoStop}%
\bibitem [{\citenamefont {Chávez~Nambo}\ \emph {et~al.}(2024)\citenamefont
  {Chávez~Nambo}, \citenamefont {Diez-Tejedor}, \citenamefont {Roque},\ and\
  \citenamefont {Sarbach}}]{Chavez_Nambo_2024}%
  \BibitemOpen
  \bibfield  {author} {\bibinfo {author} {\bibfnamefont {Emmanuel}\
  \bibnamefont {Chávez~Nambo}}, \bibinfo {author} {\bibfnamefont {Alberto}\
  \bibnamefont {Diez-Tejedor}}, \bibinfo {author} {\bibfnamefont {Armando~A.}\
  \bibnamefont {Roque}}, \ and\ \bibinfo {author} {\bibfnamefont {Olivier}\
  \bibnamefont {Sarbach}},\ }\bibfield  {title} {\enquote {\bibinfo {title}
  {Linear stability of nonrelativistic self-interacting boson stars},}\ }\href
  {\doibase 10.1103/physrevd.109.104011} {\bibfield  {journal} {\bibinfo
  {journal} {Physical Review D}\ }\textbf {\bibinfo {volume} {109}} (\bibinfo
  {year} {2024}),\ 10.1103/physrevd.109.104011}\BibitemShut {NoStop}%
\bibitem [{\citenamefont {Glennon}\ \emph
  {et~al.}(2023{\natexlab{b}})\citenamefont {Glennon}, \citenamefont
  {Mirasola}, \citenamefont {Musoke}, \citenamefont {Neyrinck},\ and\
  \citenamefont {Prescod-Weinstein}}]{Glennon:2023oqa}%
  \BibitemOpen
  \bibfield  {author} {\bibinfo {author} {\bibfnamefont {Noah}\ \bibnamefont
  {Glennon}}, \bibinfo {author} {\bibfnamefont {Anthony~E.}\ \bibnamefont
  {Mirasola}}, \bibinfo {author} {\bibfnamefont {Nathan}\ \bibnamefont
  {Musoke}}, \bibinfo {author} {\bibfnamefont {Mark~C.}\ \bibnamefont
  {Neyrinck}}, \ and\ \bibinfo {author} {\bibfnamefont {Chanda}\ \bibnamefont
  {Prescod-Weinstein}},\ }\bibfield  {title} {\enquote {\bibinfo {title}
  {{Scalar dark matter vortex stabilization with black holes}},}\ }\href
  {\doibase 10.1088/1475-7516/2023/07/004} {\bibfield  {journal} {\bibinfo
  {journal} {JCAP}\ }\textbf {\bibinfo {volume} {07}},\ \bibinfo {pages} {004}
  (\bibinfo {year} {2023}{\natexlab{b}})},\ \Eprint
  {http://arxiv.org/abs/2301.13220} {arXiv:2301.13220 [astro-ph.CO]}
  \BibitemShut {NoStop}%
\bibitem [{\citenamefont {Rodríguez~Lara}\ \emph {et~al.}(2025)\citenamefont
  {Rodríguez~Lara}, \citenamefont {Álvarez},\ and\ \citenamefont
  {Guzmán}}]{rodriguezlara2024}%
  \BibitemOpen
  \bibfield  {author} {\bibinfo {author} {\bibfnamefont {Daniela~Estefanía}\
  \bibnamefont {Rodríguez~Lara}}, \bibinfo {author} {\bibfnamefont {Iván}\
  \bibnamefont {Álvarez}}, \ and\ \bibinfo {author} {\bibfnamefont
  {Francisco}\ \bibnamefont {Guzmán}},\ }\bibfield  {title} {\enquote
  {\bibinfo {title} {Numerical solution of partial differential equations using
  the discrete fourier transform},}\ }\href {\doibase
  10.31349/revmexfise.22.020221} {\bibfield  {journal} {\bibinfo  {journal}
  {Revista Mexicana de Física E}\ }\textbf {\bibinfo {volume} {22}} (\bibinfo
  {year} {2025}),\ 10.31349/revmexfise.22.020221}\BibitemShut {NoStop}%
\bibitem [{\citenamefont {Thomas}(2013)}]{Thomas2}%
  \BibitemOpen
  \bibfield  {author} {\bibinfo {author} {\bibfnamefont {J.W.}\ \bibnamefont
  {Thomas}},\ }\href {https://books.google.com.mx/books?id=w-3SBwAAQBAJ} {\emph
  {\bibinfo {title} {Numerical Partial Differential Equations: Conservation
  Laws and Elliptic Equations}}},\ Texts in Applied Mathematics\ (\bibinfo
  {publisher} {Springer New York},\ \bibinfo {year} {2013})\BibitemShut
  {NoStop}%
\bibitem [{\citenamefont {Alvarez-Rios}\ \emph {et~al.}(2025)\citenamefont
  {Alvarez-Rios}, \citenamefont {Guzmán},\ and\ \citenamefont
  {Niemeyer}}]{AlvarezGuzmanNiemeyer_2025}%
  \BibitemOpen
  \bibfield  {author} {\bibinfo {author} {\bibfnamefont {Iván}\ \bibnamefont
  {Alvarez-Rios}}, \bibinfo {author} {\bibfnamefont {Francisco~S.}\
  \bibnamefont {Guzmán}}, \ and\ \bibinfo {author} {\bibfnamefont {Jens}\
  \bibnamefont {Niemeyer}},\ }\bibfield  {title} {\enquote {\bibinfo {title}
  {Fermion-boson stars as attractors in fuzzy dark matter and ideal gas
  dynamics},}\ }\href {\doibase 10.1103/4tkh-7hjs} {\bibfield  {journal}
  {\bibinfo  {journal} {Physical Review Letters}\ }\textbf {\bibinfo {volume}
  {135}} (\bibinfo {year} {2025}),\ 10.1103/4tkh-7hjs}\BibitemShut {NoStop}%
\bibitem [{\citenamefont {Davies}\ and\ \citenamefont
  {Mocz}(2020)}]{Moczfdmbh}%
  \BibitemOpen
  \bibfield  {author} {\bibinfo {author} {\bibfnamefont {Elliot~Y}\
  \bibnamefont {Davies}}\ and\ \bibinfo {author} {\bibfnamefont {Philip}\
  \bibnamefont {Mocz}},\ }\bibfield  {title} {\enquote {\bibinfo {title} {Fuzzy
  dark matter soliton cores around supermassive black holes},}\ }\href
  {\doibase 10.1093/mnras/staa202} {\bibfield  {journal} {\bibinfo  {journal}
  {Monthly Notices of the Royal Astronomical Society}\ }\textbf {\bibinfo
  {volume} {492}},\ \bibinfo {pages} {5721--5729} (\bibinfo {year}
  {2020})}\BibitemShut {NoStop}%
\bibitem [{\citenamefont {Lancaster}\ \emph {et~al.}(2020)\citenamefont
  {Lancaster}, \citenamefont {Giovanetti}, \citenamefont {Mocz}, \citenamefont
  {Kahn}, \citenamefont {Lisanti},\ and\ \citenamefont
  {Spergel}}]{Lancaster_2020}%
  \BibitemOpen
  \bibfield  {author} {\bibinfo {author} {\bibfnamefont {Lachlan}\ \bibnamefont
  {Lancaster}}, \bibinfo {author} {\bibfnamefont {Cara}\ \bibnamefont
  {Giovanetti}}, \bibinfo {author} {\bibfnamefont {Philip}\ \bibnamefont
  {Mocz}}, \bibinfo {author} {\bibfnamefont {Yonatan}\ \bibnamefont {Kahn}},
  \bibinfo {author} {\bibfnamefont {Mariangela}\ \bibnamefont {Lisanti}}, \
  and\ \bibinfo {author} {\bibfnamefont {David~N.}\ \bibnamefont {Spergel}},\
  }\bibfield  {title} {\enquote {\bibinfo {title} {Dynamical friction in a
  fuzzy dark matter universe},}\ }\href {\doibase
  10.1088/1475-7516/2020/01/001} {\bibfield  {journal} {\bibinfo  {journal}
  {Journal of Cosmology and Astroparticle Physics}\ }\textbf {\bibinfo {volume}
  {2020}},\ \bibinfo {pages} {001--001} (\bibinfo {year} {2020})}\BibitemShut
  {NoStop}%
\bibitem [{\citenamefont {El-Zant}\ \emph {et~al.}(2020)\citenamefont
  {El-Zant}, \citenamefont {Roupas},\ and\ \citenamefont {Silk}}]{ElZant2020}%
  \BibitemOpen
  \bibfield  {author} {\bibinfo {author} {\bibfnamefont {Amr~A}\ \bibnamefont
  {El-Zant}}, \bibinfo {author} {\bibfnamefont {Zacharias}\ \bibnamefont
  {Roupas}}, \ and\ \bibinfo {author} {\bibfnamefont {Joseph}\ \bibnamefont
  {Silk}},\ }\bibfield  {title} {\enquote {\bibinfo {title} {Ejection of
  supermassive black holes and implications for merger rates in fuzzy dark
  matter haloes},}\ }\href {\doibase 10.1093/mnras/staa2972} {\bibfield
  {journal} {\bibinfo  {journal} {Monthly Notices of the Royal Astronomical
  Society}\ }\textbf {\bibinfo {volume} {499}},\ \bibinfo {pages} {2575–2586}
  (\bibinfo {year} {2020})}\BibitemShut {NoStop}%
\bibitem [{\citenamefont {Alonso-\'Alvarez}\ \emph {et~al.}(2024)\citenamefont
  {Alonso-\'Alvarez}, \citenamefont {Cline},\ and\ \citenamefont
  {Dewar}}]{Alonso-Alvarez2024}%
  \BibitemOpen
  \bibfield  {author} {\bibinfo {author} {\bibfnamefont {Gonzalo}\ \bibnamefont
  {Alonso-\'Alvarez}}, \bibinfo {author} {\bibfnamefont {James~M.}\
  \bibnamefont {Cline}}, \ and\ \bibinfo {author} {\bibfnamefont {Caitlyn}\
  \bibnamefont {Dewar}},\ }\bibfield  {title} {\enquote {\bibinfo {title}
  {{Self-Interacting Dark Matter Solves the Final Parsec Problem of
  Supermassive Black Hole Mergers}},}\ }\href {\doibase
  10.1103/PhysRevLett.133.021401} {\bibfield  {journal} {\bibinfo  {journal}
  {Phys. Rev. Lett.}\ }\textbf {\bibinfo {volume} {133}},\ \bibinfo {pages}
  {021401} (\bibinfo {year} {2024})},\ \Eprint
  {http://arxiv.org/abs/2401.14450} {arXiv:2401.14450 [astro-ph.CO]}
  \BibitemShut {NoStop}%
\bibitem [{our()}]{ourdata}%
  \BibitemOpen
  \href {https://zenodo.org/records/19561144} {\enquote {\bibinfo {title}
  {https://zenodo.org/records/19561144},}\ }\BibitemShut {NoStop}%
\end{thebibliography}%

\appendix
\section{Regularity of solutions}
\label{app:regularity}

\subsection{Case $m=0$}

Consider the axial equation for the core configuration with $m=0$

\begin{equation}
-\frac{1}{2}\left[\partial_{r_\perp}^2\phi_0+\frac{1}{r_\perp}\partial_{r_\perp}\phi_0+\partial_z^2\phi_0\right]+V_T(r_\perp,z)\phi_0=\mu\phi_0.\nonumber
\end{equation}

\noindent The regularity conditions (\ref{eq:bcmeq0}) on the axis and equatorial plane are respectively $\partial_{r_\perp}\phi_0(0,z)=0$ and $\partial_z\phi_0(r_\perp,0)=0$. Assuming spherical symmetry of the total potential, $V_T(r_\perp,z)=V_T(r)$ with $r=\sqrt{r_\perp^2+z^2}$, the ground state inherits this symmetry so that $\phi_0(r_\perp,z)=\phi(r)$.

Using the chain rule, $\partial_{r_\perp}\phi_0=\phi'(r)\,r_\perp/r$ and $\partial_z\phi_0=\phi'(r)\,z/r$,  the regularity conditions are automatically satisfied. The Laplacian becomes $\phi''(r)+\frac{2}{r}\phi'(r)$, and the equation reduces to

\begin{equation}
-\frac{1}{2}\left[\phi''(r)+\frac{2}{r}\phi'(r)\right]+V_T(r)\phi(r)=\mu\phi(r),\nonumber
\end{equation}

\noindent or equivalently

\begin{equation}
-\frac{1}{2r^2}\frac{d}{dr}\left(r^2\frac{d\phi}{dr}\right)+V_T(r)\phi=\mu\phi.\nonumber
\end{equation}

\noindent Now, defining $u(r)=r^2\phi'(r)$ gives $\phi'(r)=u/r^2$ and $u'(r)=2r^2[V_T(r)-\mu]\phi$, the system becomes

\begin{eqnarray}
\phi'(r)=u/r^2, \nonumber \\
u'(r)=2r^2[V_T(r)-\mu]\phi. \nonumber
\end{eqnarray}

\noindent We recall that the regularity condition requires $\phi(0)=\phi_c<\infty$ and $u(0)=0$. If the potential contains a Newtonian term $V_T(r)=-\alpha/r+V(r)$ with $V(r)$ finite at the origin, then near $r=0$ one has $u'(r)=-2\alpha\phi_c r+\mathcal{O}(r^2)$, which integrates to $u(r)=-\alpha\phi_c r^2+\mathcal{O}(r^3)$. Hence $\phi'(r)=-\alpha\phi_c+\mathcal{O}(r)$ and therefore $\phi(r)=\phi_c-\alpha\phi_c r+\mathcal{O}(r^2)$, implying finally that $\lim_{r\to0}\phi(r)=\phi_c<\infty$.

\subsection{Regularity of the Axial $m>0$ Vortex Solution}

For a vortex configuration with winding number $m>0$, the axial stationary equation is

\begin{eqnarray}
&&-\frac{1}{2}\left[\frac{\partial^2 \phi_m}{\partial r_\perp ^2}
+\frac{1}{r_\perp} \frac{\partial \phi_m}{\partial r_\perp} - \frac{m^2}{r_\perp^2}\phi_m+\partial_z^2\phi_m\right] +\nonumber\\
&&~~~~~~~~~~~~~~ V_T(r_\perp,z)\phi_m=\mu\phi_m.\nonumber
\end{eqnarray}

\noindent Near the symmetry axis, the dominant contribution is the centrifugal term $-m^2\phi_m/r_\perp^2$, so to leading order the equation reduces to $\partial_{r_\perp}^2\phi_m+\frac{1}{r_\perp}\partial_{r_\perp}\phi_m-\frac{m^2}{r_\perp^2}\phi_m\simeq0$. 

Using the  ansatz $\phi_m(r_\perp,z)\sim A(z)r_\perp^p$, and substituting gives $p(p-1)r_\perp^{p-2}+pr_\perp^{p-2}-m^2r_\perp^{p-2}=0$, which implies $p^2-m^2=0$, and finally $p=\pm m$.

The branch $p=-m$ is singular at the axis and it must be discarded, which results in regular behavior $\phi_m(r_\perp,z)=A(z)r_\perp^m+\mathcal{O}(r_\perp^{m+2})$. This implies that $\lim_{r_\perp\to0}\phi_m(r_\perp,z)=0$, so that the vortex boundary condition reduces to the condition $\phi_m(0,z)=0$.

Finally, the density behaves as $\rho_m=|\phi_m|^2\sim r_\perp^{2m}$, which vanishes on the axis, $\lim_{r_\perp\to0}\rho_m=0$.

\end{document}